\RequirePackage{fix-cm}
\documentclass[smallextended]{svjour3}       
\smartqed  

\usepackage{nicematrix}
   \usepackage{threeparttable}
\usepackage{graphicx}
\RequirePackage{fix-cm}
\smartqed  

\usepackage{amssymb}
\usepackage{graphicx}
\usepackage{todonotes}
\usepackage[most]{tcolorbox}

\usepackage{hyperref}

\usepackage{float}
\usepackage{xcolor}
\usepackage{diagbox}
\usepackage{epstopdf}
\usepackage{caption}
\usepackage{subfig}

\usepackage{listings}
\usepackage{xcolor}

\definecolor{codegreen}{rgb}{0,0.6,0}
\definecolor{codegray}{rgb}{0.5,0.5,0.5}
\definecolor{codepurple}{rgb}{0.58,0,0.82}
\definecolor{backcolour}{rgb}{0.95,0.95,0.92}

\lstdefinestyle{mystyle}{
 backgroundcolor=\color{backcolour},  commentstyle=\color{codegreen},
 keywordstyle=\color{magenta},
 numberstyle=\tiny\color{codegray},
 stringstyle=\color{codepurple},
 basicstyle=\ttfamily\footnotesize,
 breakatwhitespace=false,     
 breaklines=true,         
 captionpos=b,          
 keepspaces=true,         
 numbers=left,          
 numbersep=5pt,         
 showspaces=false,        
 showstringspaces=false,
 showtabs=false,         
 tabsize=2
}

\begin{document}

\title{Dissecting Smart Contract Languages: A Survey 
}


\author{Majd Soud        \and
        Gísli Hjálmtýsson        \and
        Mohammad Hamdaqa 
}


\institute{M.Soud 
             \\
              Department of Computer Science, Reykjavik\\ University, Reykjavik, Iceland\\
              \email{majd18@ru.is}  \\          
           \and G.Hjálmtýsson \\
           University, Reykjavik, Iceland\\
                 \email{gisli@ru.is}     \\ \and
                M.Hamdaqa\\
              \email{mhamdaqa@polymtl.ca }  
}

\date{Received: date / Accepted: date}

\maketitle

\begin{abstract}
Blockchain is a distributed ledger technology that gained popularity for enabling the transformation of cryptocurrency between peers without mediation by a centralized third-party authority. Smart contracts expand the applications of blockchain technology and have played a role in its widespread adoption. Smart contracts are immutable digital programs that are deployed on blockchains to codify agreements between parties. Existing smart contract implementations have faced challenges, including security vulnerabilities, leading to significant losses and concerns. This has stimulated a wave of attempts to improve Smart Contract Languages (SCLs) to overcome implementation challenges and ensure code quality, producing many languages with diverse features. Scholars have made some attempts to classify SCLs and clarify the process of selecting an SCL, but to the best of our knowledge, no comprehensive survey of existing SCLs has been published. Our work surpasses earlier efforts by evaluating a considerably larger set of SCLs, in greater depth, to ease the process of SCL selection for blockchain research and implementation. In this paper, we (1) propose a robust framework for comparing existing SCLs, (2) analyze and discuss 36 SCLs, addressing issues beyond those used to construct the comparison framework, and (3) define new parameters for future research and development of SCLs. The survey provides a guide for those who intend to select or use an SCL to implement smart contracts, develop new SCLs, or add new extensions to the existing SCLs.
 
\end{abstract}

\section{Introduction}
\label{intro}
Blockchain technology has gained remarkable attention from many research and industry experts. By some estimates, blockchain technology has been used to manage transactions and assets with a total estimated value of over two hundred billion dollars in the past ten years. A blockchain is a decentralized and immutable digital ledger that records transactions across a network of computers. It supplants traditional trust models that rely on third parties, and authorities, such as banks, to guarantee the integrity of currencies and transactions. In its place, new peer-to-peer mechanisms that use cryptography and distributed consensus are used to make transactions verifiable~\cite{pilkington2016blockchain}. Early blockchain applications focused specifically on cryptocurrencies. Smart contracts to codify more elaborate transactions greatly increase the range of applications. Blockchain technology continues to find new uses today, as new coding protocols and agreements are implemented in the form of smart contracts.
The evolution of smart contracts ushered in an era called “Blockchain 2.0”.  Smart contracts are computer programs that serve as an executable, self-enforcing, verifiable agreement~\cite{oliva2020exploratory}. Their inherent credibility effectively eliminates the need for third-party authorities to enforce contractual terms. 

The term “smart contract” was originally coined by Nick Szabo in~\cite{szabo1996smart}, to refer to software or protocols that facilitate all steps of the contracting process, from searching for contracts to negotiation, commitment, and adjudication.
Smart contracts have become widely used for two reasons. First, there have been advances in the technology to create smart contracts. Second, awareness of their benefits is spreading. Smart contracts are used in many domains where the consequences of errors are severe. This includes e-commerce applications~\cite{e1}, financial technology~\cite{f1}, health care and life science applications~\cite{h1} and many others.

In these critical domains, poorly designed and implemented smart contracts can cause significant harm and lead to large financial losses. This happened, for example, in the infamous DAO incident~\cite{DAO2} and in the Parity multi-sig incidents~\cite{parity1}. In the DAO case, an attacker exploited a reentrancy vulnerability in the code of a smart contract and gained control of \$50M. In 2018, an attacker exploited an integer overflow vulnerability in the contract underpinning the “Proof of Weak Hands” contract, making off with 866 Ether worth \$2.3M. More generally, it is estimated that the combined value held by Ethereum smart contracts deployed on the public blockchain by the end of 2019 was worth more than eighteen billion dollars. We observe that these numbers inherently make the underlying smart contracts attack targets, and highlight the risk of poorly designed and executed smart contracts. To protect this kind of value, it is crucial that smart contracts are implemented correctly and free of code vulnerabilities.

It is challenging to assess the security and soundness of smart contracts. Developers have to consider the lack of execution control and the immutable characteristics of smart contracts once they are deployed to the blockchain. They are also responsible for rigorously monitoring smart contract functionality and run-time behavior in an environment of continual platform updates and new potential security risks. A number of Smart Contract Languages (SCLs) have emerged as a response to these challenges. SCLs help developers (1) implement correct, safe, and secure smart contracts, (2) deploy the smart contracts across diverse blockchain platforms and (3) satisfy the specific functional requirements in different application domains. A large and heterogeneous set of SCLs now exists, many with overlapping goals and features. Clarification is needed to help developers choose the SCL that best fits their application requirements, and stability and security needs. 
While there have been previous attempts to survey Smart Contract Languages (SCLs), such prior work has reviewed relatively few languages and analyzed relatively few of their features.~\cite{parizit} have analyzed three domain-specific programming languages: Solidity, Pact, and Liquidity, focusing exclusively on usability and security and leaving out other features of SCLs.~\cite{harz} have conducted a broader review surveying 18 SCLs, focusing on security features, as well as aspects of their language programming paradigm, level of abstraction, instruction set supported, language semantics, and support for metering.

Many currently used SCLs could not be reviewed in these studies because they emerged around or shortly after the time these surveys were conducted. The ongoing proliferation of SCLs remains a challenge in this field. It takes substantial effort to closely analyze many important characteristics and features of numerous languages.
In our paper, we embrace this challenge, systemically surveying 36 existing SCLs, with a focus on a broad set of their characteristics and features. Our work is influenced by the concerns mentioned above and by additional insights that arise in the course of our work. We illustrate how SCLs differ in their scope and goals. We also classify them and present a framework for comparing characteristics across languages.
\\
\textbf{Contribution.} The overall contributions of this paper are as follows: 
\begin{itemize}
   \item 	We introduce a comprehensive framework designed for the comparative analysis of Smart Contract Languages (SCLs), which we employ to thoroughly investigate and evaluate the SCLs currently in widespread use today.
   
  \item 	We present a detailed analysis and engage in extensive discussions concerning 36 SCLs, addressing a range of issues that extend beyond the scope of the initial comparison framework.
  
  \item 	 We put forward novel parameters that hold the potential to shape the direction of future research and development in the realm of Smart Contract Languages (SCLs).
\end{itemize}
Significantly, our work contributes to the field by conducting a thorough analysis within a thoughtfully structured comparative framework, thus adding depth to the existing body of research.

\textbf{Paper Organization.} Section 2 summarizes existing survey articles and literature reviews on SCLs. Section 3 illustrates the methodology and research process we used to conduct this survey. Section 4 presents the classification and comparison framework we used in our own SCL survey. In Section 5, we answer the defined Research Questions (RQs). In Section 6, we highlight our main findings. In Section 7, we discuss some future directions for research and development. Finally, we conclude and discuss some of the limitations of our work in Section 8.
\section{Related Work}
This section briefly discusses and summarizes previous survey articles that compare SCLs and empirically analyzes their characteristics or some of their characteristics. In Table 1, we tabulate all the surveys discussed in this section.

Although smart contracts are increasingly gaining interest, few papers concentrate on SCLs such as Bartoletti et al.~\cite{bartoletti}, which conducted an empirical study of smart contracts applications, platforms, and design patterns. They looked at six current, publicly accessible blockchain platforms that were supported by communities of developers; namely, Bitcoin~\cite{bitcoin}, Counterparty~\cite{counterparty}, Ethereum~\cite{ethereum}, Lisk~\cite{lisk}, and Stellar~\cite{stellar}. They assessed three qualities of these platforms. First, they indicated if the platform had its own unique blockchain, or if it depended on another one. Secondly, they looked at consensus protocols in public blockchains, and if private blockchains are supported in the platforms or not. Finally, they named the SCLs that were used to implement the smart contracts deployed on these platforms. Our work exceeds the scope of this earlier study both in the number of SCLs examined and the number of SCL characteristics discussed.

Parizi et al.~\cite{parizit} conducted an evaluation of three domain-specific SCLs in terms of usability and security. They quantified the usability of Pact, Solidity, and Liquidity by conducting an experiment on 15 test subjects, asking them to implement some smart contracts using these three languages. They then used a built-in timer to measure the time the test subjects spent implementing the contract's code for each SCLs. SCL security was measured by executing the implemented contracts and using available security analysis tools to probe for common security vulnerabilities, such as Timestamp Dependency and Assertion Failure. The authors indicate that SCL research is almost new and describe their work as an early contribution to the field of SCL studies. They also offered suggestions for improving SCL comparisons and generating stronger research results, such as involving more test subjects in the experiments, running more test contracts with several context parameters, and including new upcoming SCLs. These suggestions are incorporated into our work.

Harz et al.~\cite{harz} focused on smart contract implementation problems that result in sharp losses for end users. They surveyed 18 SCLs, emphasizing security features and providing an overview of each SCL that covered its instruction set, semantics, metering, and paradigm. The commonly used verification methods and tools for smart contracts were reviewed, detailing their verification methods, supported SCLs, and automation levels.~\cite{anderson2016new} compared the usage patterns and transactions of Ethereum, Peercoin~\cite{king2012peercoin} and Namecoin~\cite{namecoin}. Areas of negligence in smart contract design for Ethereum were discussed. For Namecoin, they analyzed the use of name registration and its development over time. 
A similar overview of several contract platforms was offered by Seijas et al.~\cite{seijas2016scripting} a brief description of the three scripting languages for Ethereum, Nxt, and Bitcoin was presented in the paper. 

\begin{table}
\caption{Summary of related surveys and research papers.}
\label{tab:10}       
\begin{tabular}{p{1.3cm}p{2.9cm}p{2cm}p{2cm}p{1.5cm}}
\hline\noalign{\smallskip}
Research Reference & Surveyed SCLs  & Research Focus & Discussed Language Features & Research Approach  \\
\noalign{\smallskip}\hline\noalign{\smallskip}
~\cite{bartoletti} & Bitcoin script, solidity, JavaScript (Lisk SDK) & Smart contracts platforms and design patterns & No discussion on language features & Empirical Study \\

~\cite{parizit} & Pact, Solidity, and Liquidity & Language usability and security & Common security vulnerabilities per language & Experimental Study\\

~\cite{harz} & Solidity, Vyper, Bamboo, Flint, Pyramid, Obsidian, Rholang, Liquidity, DAML, Pact, Simplicity, Scilla, EthIR, IELE, Yul, EVM, eWASM, Michelson & Safety of the smart contracts & Instruction set, semantic, metering, and paradigm, verification methods, and automation level. & Comparative Study \\
\noalign{\smallskip}\hline
\end{tabular}
\end{table}

Although we are more than a decade after the first blockchain was proposed and blockchain technology is widely used, SCLs are still poorly analyzed. Prior studies essentially review and compare a small set of blockchain platform features discuss SCLs as a secondary topic and leave a gap in the empirical knowledge about the available SCLs. In our work, we provide the reader with a comprehensive survey in which we classify and compare the 36 existing SCLs. We provide smart contract developers and users with a classification framework to compare these languages. We also discuss the pragmatics, targets, and scope of these languages. Then, we discuss the characteristics and features that are relevant to language engineers, and other characteristics that help regular users. This survey is intended to help those who intend to implement smart contracts, select the appropriate SCL, design a new or extend an existing SCL, or conduct research in the field of SCLs.

 
\section{Research Review Process}
The study of blockchain and smart contract technology is highly influenced by industrial research. Consequently, we have conducted a Multivocal Literature Review (MLR) to list and review the available SCLs in both industrial and academic fields. MLR is a form of Systematic Literature Review (SLR) that involves both “gray” and “white” literature~\cite{ogawa1991towards}. The 'gray' literature consists of informal documents such as web pages, videos, and blogs. By contrast, “white” literature refers to formal research published in conference proceedings and academic journals. We start by planning the MLR and defining the survey goal and set of questions. We attempt to answer these questions by conducting the research described in Section 3.1. Then, we define our data collection process and our inclusion and exclusion criteria in Section 3.2. The process of data extraction and structuring is discussed in Section 3.3. Finally, the results are analyzed based on the comparison framework that we introduce in Section 4.

\subsection{Planning the Review }
This survey's primary goal is to provide an overview of the available SCLs, understand the current state of the art in the development of SCLs, classify their common characteristics, and identify the gaps and future research parameters for SCLs. In particular, we identified the following research questions (RQs):  
\begin{itemize}
  
  \item 	\textbf{RQ1:  What are the prevalent contemporary Smart Contract Languages (SCLs) in use today?}
  \item 	\textbf{RQ2:  What are the primary objectives and purposes associated with each of the examined SCLs?}
  \item \textbf{RQ3: From a language engineering perspective, what are the distinctive characteristics of each SCL?}

\end{itemize}

Our examination of these questions suggests directions for future research regarding the development of SCLs. The survey does not claim to cover all SCLs developed to date. This is infeasible given the rapid rate at which new languages are developed in this domain. However, the survey covers the most common SCLs known at the time of submitting this survey. We base our work on available information, and this limits the resources available to us about each covered language.  

\subsection{Data Collection Process}

Our data collection process consists of four refinement steps. We start with a generic keyword-based search with keywords included from the defined RQs, and then we filter the results by focusing on specific queries and applying our inclusion and exclusion criteria. After that, we prune the results by removing duplicates and irrelevant sources.
In this work, we decided to search for all data resources that have the term “smart contract” for the period from 1994, when the smart contract concept was first proposed, to January 2021. Due to the highly diversified nature of the topic, and the lack of academic literature that covers SCLs, we used the MLR approach as mentioned before. 
Based on the research questions proposed in Section 3.1, we defined initial search keywords and queries. We considered the terms “smart contract”, “blockchain”, “language”, “platforms”, “patterns”, and “domain” and their combinations as search keywords to query various data sources from the “gray” and “white” literature. For the white literature, we limited the search to articles that are indexed by the search keyword 'smart contract'. We searched the following scholarly search engines as shown in Table 2: ACM Digital Library\footnote{http://portal.acm.org}, Science Direct\footnote{http://www.sciencedirect.com}, and SpringerLink \footnote{http://www.springerlink.com}. Then, we selected the queries with the highest recall. For the gray literature, we applied these keywords to query Google search engine.  
In order to limit the scope of our review, we only considered sources that proposed a smart contract language and sources that describe SCL concepts. To limit the scope of our review, we only considered sources that proposed a smart contract language and sources that describe SCL concepts. Moreover, to improve confidence in covering all related studies, a snowballing procedure approach~\cite{wohlin2014guidelines} was applied. Also, we did backward and forward snowballing. In the backward snowballing, we examined the reference list for each study and removed any study that did not match this survey's proposed criteria. In the forward snowballing, we reviewed the studies depending on their citations. Thus, we analyzed any study that cites a specific study. Then, we used the following inclusion and exclusion criteria to filter the relevant studies and sources:\\
\textbf{Inclusion criteria}
\begin{itemize}

    \item 	Data sources that propose SCLs. 
    \item 	Data sources that include language concepts about SCLs.
    \item 	Data sources that compare several SCLs.
    
\end{itemize}
\textbf{Exclusion criteria}
\begin{itemize}
 \item 	Data sources that solely describe Blockchain platforms without referring to smart contracts.
 \item 	Data sources that are not written in English.
\end{itemize}
After filtering the results, the result was a list of relevant data sources. However, this data set contained duplicate sources as well as duplicate content within sources. To remove duplicates, we applied two pruning stages.

\begin{table}

\caption{Summary of results of search queries after pruning stage 1.}
\label{tab:12}       
\begin{tabular}{p{2 cm}p{7 cm}p{1.5cm }}
\hline\noalign{\smallskip}
Electronic Database & Search Query & \# of studies  \\
\noalign{\smallskip}\hline\noalign{\smallskip}
ACM Digital Library & "query": {(blockchain, smart, contracts, language, smart contract, platform, framework) AND keywords. author.keyword:(+"smart contract")} "filter":{"publication Year":{"gte":1994 -2020}} & 98 \\
ScienceDirect & (blockchain OR contract OR language OR design OR pattern OR platform) and KEYWORDS(“smart contract")[All Sources(Computer Science)] [All Sources(Computer Law \& Security Review)] & 191 \\
SpringerLink & "smart contracts" AND (smart OR contract OR blockchain OR language OR model OR platform) AND 1994-2019 & 111 \\
Total \# &   & 400\\
\noalign{\smallskip}\hline
\end{tabular}
\end{table}

\textbf{Pruning stage 1:} We extracted the relevant data from the selected data sources. Then, we removed all duplicate sources. After that, we identified duplicate sources by taking into account the title, introduction, authors, conclusion, and the year of the publication. At the end of this stage, we had 400 unique articles and 150 web resources resulting from Google searches, as shown in Table 2 and Table 3.\\

\textbf{Pruning stage 2:} We performed manual selection based on title and content, for gray and white sources. Since it is difficult to determine if the data source is relevant or not using only the title, at this stage, we reviewed the content of each of the selected sources to ensure that it covered all the required data needed for our comparison. This left us with 60 articles and 90 web-based sources. Some were surveys related to SCLs, and others were proposed SCLs. Table 3 shows a summary of the total number of data sources at each stage of the selection process.

\begin{table}

\caption{A summary of this data collection research process.}
\label{tab:13}       
\begin{tabular}{p{3.5 cm}p{3 cm}p{3cm }}
\hline\noalign{\smallskip}
Stage & \# of white literature Articles  & \# of gray literature results \\
\noalign{\smallskip}\hline\noalign{\smallskip}
Step1: General Search &  12,500 & 14,700,000 \\
Step2: Query Search and Filtering & 5000 &108,000 \\
Step3: Pruning stage 1 & 400 & 150 \\
Step: Pruning stage 2 &  60 &  90 \\ \hline
Resulted SCLs &   15 & 20 \\
Total \# SCLs from both gray and white literature & 36 & \\
\noalign{\smallskip}\hline
\end{tabular}
\end{table}
As SCLs have evolved remarkably in recent years, recent work was prioritized. The main sources involved in the final data pool were classified as follows:
\begin{itemize}
    
\item  Academic literature related to SCLs and blockchain platforms.
\item 	Official documentation of current and available SCLs.
\item 	All GitHub repositories that belong to SCLs. 
\item 	Discussion forums and internet blogs related to SCLs.

\end{itemize}

After filtering these sources, the final number of SCLs covered was 36 languages, including the SCLs that were discussed in the related work.

\subsection{Data Extraction }
Our goal was to extract relevant data from our sources and compare them systematically. First, we manually inspected the data sources. This formed the basis for our proposed classification framework, described in Section 4. The framework was used to create a database where each column refers to a property of a class in the proposed framework. We used the data sources to populate the database, which we then used to compare the different languages.  

\section{Proposed Framework}
In this section, we introduce a framework for comparing and classifying the available SCLs. Establishing such a framework was based on a thorough investigation of each of the SCLs’ scope, characteristics, official documentation, websites, and published papers. Our framework is shown in Figure 2, in which the main classes and their potential manifestations are provided. These classes and attributes were used to classify and compare the 36 SCLs.

\begin{figure*}
  \includegraphics[width=1.05\textwidth]{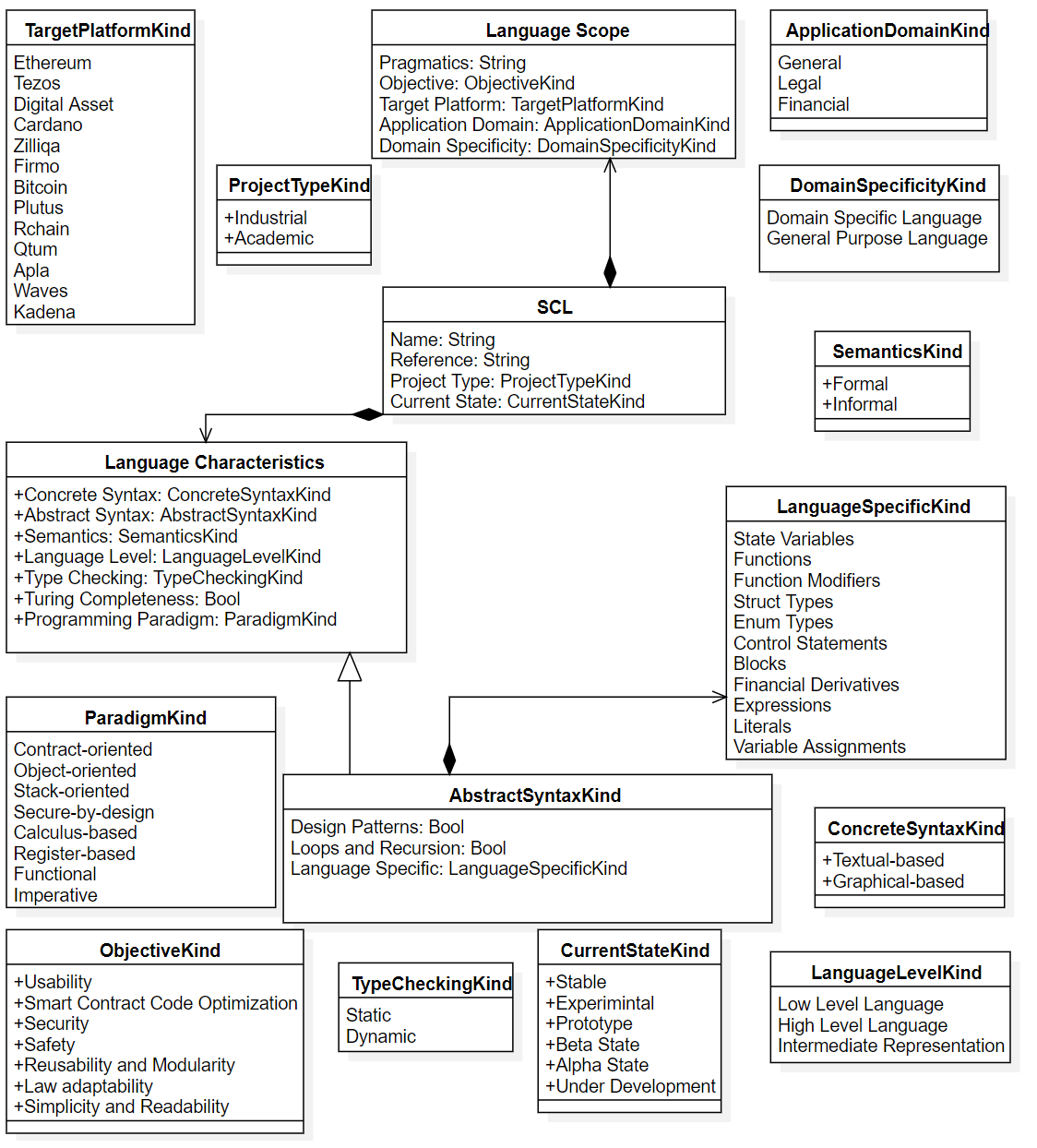}
\caption{Our proposed comparison framework.}       
\end{figure*}

\subsection{Language scope}
The smart contract language scope consists of the language pragmatics and objectives, the target blockchain platform, the application domain of the language, as well as whether the language is generic or domain-specific. 
\subsubsection{Pragmatics and Objectives }
SCL pragmatics refers to the purpose behind creating the language. It covers the SCL’s objectives (O), as stated by the creators, beyond the definition of the language itself. Pragmatic objectives are usually declared in the promotional content for a language and should not be used as a reliable fact sheet. However, pragmatic objectives strongly influence aspects of an SCL’s design, such as its syntax and semantics. While language characteristics could be used to crosscheck pragmatics, to avoid subjectivity in this work, we took the pragmatic claims made for each of the 36 SCLs as a parameter for comparing languages.

\begin{table}

\caption{Usage frequency of the proposed objective categories among SCLs.}
\label{tab:14}       
\begin{tabular}{p{4.5 cm}p{4 cm}}
\hline\noalign{\smallskip}
Objective categories & \# of SCLs that used it to describe their objectives \\
\noalign{\smallskip}\hline\noalign{\smallskip}
O1 - Usability & 25 \\
O2 - Security & 4 \\
O3 - Safety & 12 \\
O4 - Smart Contract Code & 3\\ 
O5 - Reusability and Modularity & 7  \\
O6 - Law Adaptability & 1 \\

\noalign{\smallskip}\hline
\end{tabular}
\end{table}

These claims about the objectives (O) of an SCL generally fall into six categories:\\
\textbf{O1} - Usability: The usability of an SCL is the degree to which it can be used by the intended users (developers, lawyers, contract participants) in an easy and effective way to achieve their intended objectives and satisfaction.\\
\textbf{O2} - Security: Implementing a secure smart contract that avoids security vulnerabilities is a common objective of many SCLs. Languages that proclaim security as one of their objectives normally support verification and validation to ensure that behavior state translations, for any defined entity in the smart contract, are the same as specified in the smart contract code, and nothing from outside the smart contract code, embodying security coding patterns and supporting security analysis.\\ 
\textbf{O3} - Safety: A safe SCL can guarantee that the runtime of a smart contract cannot be subverted as a result of programming mistakes.\\
\textbf{O4} - Smart Contract Code Optimization: Another key objective that appears as part of the pragmatics of some SCLs is the ability to optimize the code to run on a target platform. The goal is normally to improve quality and efficiency when running the smart contract by reducing code size, managing memory, or reducing the cost of execution on the target blockchain platform (e.g., if the smart contract uses more gas than necessary in Ethereum).\\
\textbf{O5} - Reusability and Modularity: SCLs that focus on reusability and modularity separate the functionality of a smart contract into independent units so that each unit can be used to achieve a specific action and can be readily reused by the developer to achieve that action.\\
\textbf{O6} - Law Adaptability: SCLs that adapt to law regulations by providing law programming directives that can be used in the smart contract.
Table 4 indicates how many SCLs claimed each above-listed objective as a goal or benefit of their design. 
\subsubsection{Target Platform}
The target platform refers to the blockchain platform where a smart contract is deployed, executed, and managed. Some SCLs are designed to target one or more blockchain platforms.  Part of analyzing the scope is to explore the target platforms of the SCLs we reviewed.

\subsubsection{Application Domain}
We separated SCL application domains into three classes: (1) general – which means the language can be used to implement any contract type, (2) financial – which refers to contracts that only include financial directives (3) non-financial legal contracts – legally binding contracts that do not include financial directives.

\subsubsection{	Domain specificity}
We defined two domains for classifying an SCL. It can either be a General-Purpose Language (GPL) or a Domain Specific Language (DSL). DSLs are specifically designed for programming smart contracts alone. GPLs are languages that can be used to create several types of programs, including smart contracts.  

\subsection{Language Characteristics}
The smart contract language characteristics focus on the features and properties of the language. In this survey, we explore the syntax, semantics, level of abstraction, and programming paradigm, in addition to the language support of type checking and whether it is Turing complete or not.

\subsubsection{Syntax}
The abstract syntax of SCLs represents the concepts (or constructs) underlying the language and how these concepts/constructs are linked and related to each other. Every SCL has a concrete syntax and an abstract syntax. Both are essential parts of the language definition. In this work, we reviewed both the concrete and abstract syntax of all 36 languages. We divided concrete syntax into two categories: text-based and graphical. We found that concrete syntax often mirrored the constructs of popular GPLs, so we included that in our comparison. For abstract syntax, we examined three main abstract syntax attributes in our classification scheme related to programming and structuring smart contracts; those are design patterns, loop and recursion support, and language-specific constructs.

\begin{itemize}
    \item	\textbf{Design patterns}\\
   Design patterns describe elegant solutions to specific problems. An SCL either can have predefined code patterns that developers can use without the need to implement them from scratch, or it does not provide predefined code patterns. Predefined code patterns are reusable code idioms (e.g., functions, classes) that are implemented and ready-to-use solutions. For example, a language might support security code patterns (e.g., control access code pattern) and provide predefined idioms to address some of the common security pitfalls.
    \item\textbf{	Loop and Recursion support}

Our classification scheme singled out loop and recursion programming structures as particularly important for SCL classification, since they have been exploited multiple times in smart contracts attacks like reentrancy attacks~\cite{atzei2017survey}. In our classification, an SCL can either support looping and recursion or not.  
     \item	\textbf{Language specific constructs}\\
An SCL provides developers with a specific set of programming components that they can use in order to implement smart contract functionality. These programming components differ from one SCL to another. In this paper, we summarize all the programming components each SCL allows. 
    
\end{itemize}

\subsubsection{Semantics}
The semantics of any programming language give meaning to syntactically correct components and expressions~\cite{briscoe2011introduction}. SCL semantics generally involves transforming the correct syntactic components into a determined semantic domain. In this work, we classified semantics as either formal or informal. Formal semantics defines the program's behavior in a well-formed, unambiguous manner; unfortunately, they are typically harder to use and understand by the novice programmer. On the other hand, in informal semantics, the program requires interpretation, where compilers define the program's behavior. 

\subsubsection{Language level}
SCLs can be classified into three different levels: high-level languages, Intermediate Representation (IR) languages, and low-level languages. A high-level language is close to human language and is easily understood. A low-level language is a language with no abstraction, and its functions are very close to the processor instructions.  An IR language is intermediate between high-level and low-level languages and is mostly used to implement optimized programs and/or reasoning about source code properties such as code safety. 

\subsubsection{Type checking}
Checking and enforcing type constraints is known as type checking. It can be classified as static or dynamic. Static type checking is performed at compile-time. Dynamic type checking takes place at run-time. Some languages combine them both.

\subsubsection{Turing completeness}
One profound decision in the field of blockchain design is the property of Turing completeness. A programming language is Turing complete if it has the ability to simulate any single-tape Turing machine. In other words, it has a finite set of states, a defined way to change state, and reusable storage for reading and writing operations. We classify SCLs as either Turing complete or Turing incomplete. Moreover, we do not consider the role of the execution time environment in limiting or expanding the functionality of the language.    
\subsubsection{Programming paradigm}
In the course of our analysis, we observed that SCL designers describe programming paradigms in their documentation. We collected these descriptions and classified them as follows.
\begin{itemize}
 \item \textbf{Contract-oriented: } Languages that focus on client-supplier relationships between software modules by formally and precisely specifying the interfaces between those models.
 \item	\textbf{Calculus-based:}  Formal languages that use logical expressions and mathematical proofs to express computations and business logic. 
 \item\textbf{Functional programming:} A declarative programming paradigm that expresses the program logic and computations as pure functions and avoids mutable data. 
 \item\textbf{Imperative programming:} The program is expressed as a sequence of statements that specify how to reach specific goals. An algorithmic sequence of commands and instructions is needed to update the contract's state in explicit steps.
 \item\textbf{Secure by design:} Languages designed with significant effort applied to provide constructs that enable programmers to write programs that are free of vulnerabilities and bugs.  
 \item\textbf{Object-oriented programming:} Languages that decompose applications into objects that encapsulate data fields with the procedures that work on these data fields.
 \item	\textbf{Stack-oriented:} Languages based on a stack machine model for passing parameters. A stack-based language can have more than one stack, and each stack might serve a different purpose.  

\end{itemize}

\begin{table}

\caption{A summary of available SCLs.}
\label{tab:15}       
\begin{tabular}{p{2 cm}p{2.7 cm}p{2cm }p{3cm }}
\hline\noalign{\smallskip}
SCL Name & References  & Project Type & Current State  \\
\noalign{\smallskip}\hline\noalign{\smallskip}
ADICO &~\cite{adico} & Academic & Prototype\\
Babbage &~\cite{babbage}  & Not specified & Prototype\\
Bamboo &~\cite{Pirapira}  & Industrial & Alpha state\\
Bitcoin Script &~\cite{Script} & Industrial & Stable\\
BitML &~\cite{atzei2019developing} & Academic & Under development\\
DAML &~\cite{daml} & Industrial & Stable \\
E &~\cite{ERights} & Academic & Stable\\
eWASM &~\cite{ewasm} & Industrial & Under development\\
Fi &~\cite{goodman2014tezos} & Industrial & Under development\\
Findel &~\cite{biryukov2017findel} & Academic & Under development\\
FirmoLang &~\cite{FirmoLang} & Industrial & Under development\\
Flint &~\cite{schrans2018writing} & Industrial & Alpha state\\
Formality &~\cite{moonad} & Industrial & Under development\\
Functional solidity &~\cite{Raineorshine} & Not specified & Prototype\\
IELE &~\cite{kasampalis2019iele} & Academic & Prototype\\
Ivy &~\cite{Ivy-Lang} & Academic & Prototype\\
L4  &~\cite{Legalese} & Academic & Alpha state\\
Liquidity &~\cite{OCamlPro} & Industrial & Under development\\
LLL &~\cite{LLL} & Industrial & Under development\\
Logikon &~\cite{Logikon} & Industrial & Experimental\\
Lolisa &~\cite{yang2018lolisa} & Academic & Under development\\
Michelson &~\cite{goodmanimichelson} & Industrial & Under development\\
Obsidian &~\cite{coblenz2017obsidian}& Academic & Stable\\
Plutus   &~\cite{Plutus} & Both & Under development\\
Pyramid &~\cite{Pyramid} & Industrial & Stable\\
QSCL  &~\cite{Qtum} & Industrial & Stable\\
Rholang  &~\cite{eykholt2017rchain} & Industrial & Under development\\
RIDE &~\cite{Waves}  & Both & Under development\\
Scilla &~\cite{sergey2018scilla} & Academic & Under development\\
Serpent &~\cite{delmolino2015programmer} & Industrial & Stable\\
Simplicity &~\cite{o2017simplicity} & Academic & Under development\\
Simvolio &~\cite{Simvolio} & Industrial & Prototype\\
Solidity &~\cite{solidity,modi2018solidity,Solidity2} & Industrial  & Stable\\
SolidityX &~\cite{Solidityx} & Industrial & Beta state\\
Viper/Vyper &~\cite{Viper} & Industrial & Beta state\\
Yul &~\cite{Yul} & Industrial & Under development\\

\noalign{\smallskip}\hline
\end{tabular}
\end{table}
\section{Survey Results and Discussion}
In this section, we present a full comparison between the 36 SCLs we investigated, using our proposed framework. This section is organized to address the research questions (RQs) we introduced in Section 3.1.
\subsection{The Smart Contract Languages Surveyed}

\begin{figure*}

  \includegraphics[width=0.95\textwidth]{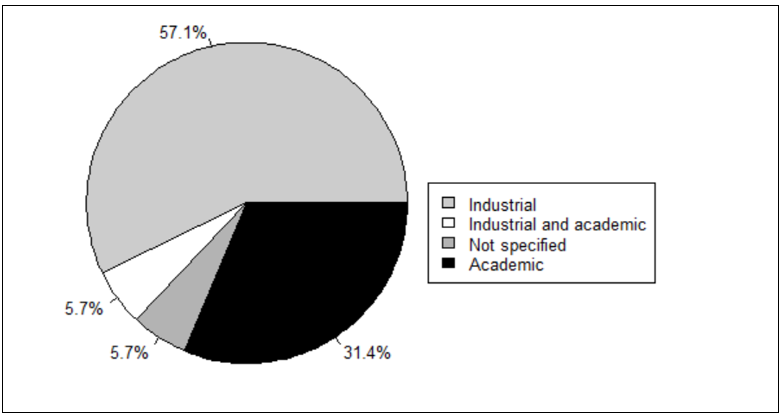}

\caption{The distribution of surveyed SCLs based on whether they are created as part of academic or industry based projects.}
\label{fig:3}       
\end{figure*}
\begin{figure*}

  \includegraphics[width=0.95\textwidth]{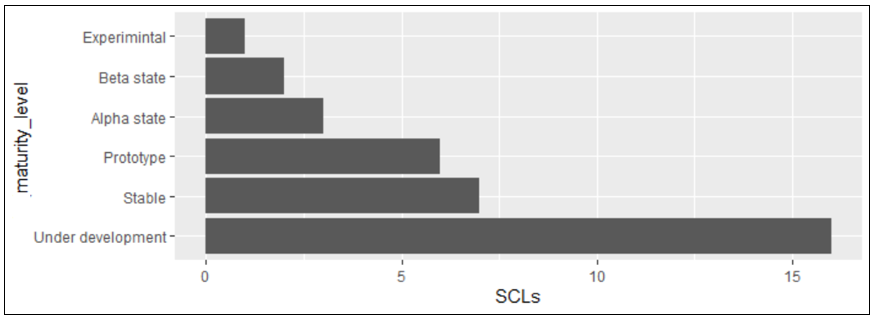}

\caption{The maturity levels of the surveyed SCLs.}
\label{fig:4}       
\end{figure*}

To answer the question of what are the common existing SCLs (RQ1), we have tabulated the list of the 36 SCL we surveyed in Table 5, classified by type of project (academic, industry), and the maturity level of the project (i.e., its current development state). As shown in Figure 3, more than half of these surveyed languages are industry-based projects. Eleven out of 36 SCLs (i.e., around 31\%) were created or defined as academic research projects. Only about 6\% of the surveyed languages were collaborations between industry and academia.

We also classified the surveyed SCLs based on their maturity level and current state. A stable language is a language that is being actively used to create smart contracts, and that is at high maturity levels. We found 8 stable SCLs: Solidity, Obsidian, Pyramid, Digital Asset Modeling Language (DAML), Bitcoin Script, E, Qtum Smart Contract Language QSCL, and Serpent. Alpha state languages included Bamboo, L4, and Flint. Alpha is the first stage of testing a language, and it is the first phase in the language development life cycle. Viper and SolidityX are in the beta stage, which is the second phase in the development of a programming language. Six languages are prototype languages: Functional solidity, Babbage, Intermediate-Level Blockchain Language (IELE), Ivy, Simvolio, and ADICO. A prototype language provides a model that allows developers to work on object cloning and prototyping. Logikon is an experimental language. The rest of SCLs are still under development (see Figure 4).

To sum up, we reviewed the available SCLs in this section, their target fields (industry or academy), and their maturity levels. We found that most of these SCLs target the industrial field. As well as most of these SCLs are still in the development phase. Only 7 out of 36 SCLs are stable by the time of writing this survey.  

\begin{table}

\caption{A comparison of the scope of SCLs.}
\label{tab:16}       
\begin{tabular}{p{1.2 cm}p{4 cm}p{1cm }p{1cm }p{1.2 cm }p{1cm }}
\hline\noalign{\smallskip}
SCL Name & Pragmatics  & Objective & Target Platform & Application Domain & Domain specificity \\
\noalign{\smallskip}\hline\noalign{\smallskip}
ADICO & Creates mapping methods that simplify the generation of Smart Contracts (SCs). & O1 & No project & General & DSL\\
Babbage & Creates SCs using a visual mechanical metaphor. & O1 & Ethereum & General & DSL\\
Bamboo & Implements SCs improving reentrancy behavior and formal verification.  & O2 & Ethereum & General & DSL\\
Bitcoin Script& Allow users to specify certain conditions (contracts) that determine how spending Bitcoins shall be done and who shall have access to the Bitcoins being transferred in the network& O1 & Bitcoin & General & DSL \\
BitML & Creates SCs that regulate transfer of bitcoins among participants. & O1 & Bitcoin & General & DSL\\
DAML & Implements distributed ledgers and smart contracts, enables high-level specification of real-time business logic. & O3  & Digital Asset  & Financial & DSL\\
E & Object modelling of secure SCs and distributed persistent computation. & O5 & No project  & General & DSL\\
eWASM & Generates SCs with accessible development for users. & O1 & Ethereum  & General & DSL\\
Fi & Creates SCs that compile directly to valid and well-typed Michelson code. & O5 & Tezos & General & DSL\\
Findel & Models financial derivatives on blockchains and SCs. & O1 & Ethereum  & Financial & DSL\\
FirmoLang & Creates verified financial contracts with a diminishing margin of error in automated SCs & O3 & FIRMO & General & DSL\\
Flint & Creates robust SCs with contract safety in mind. & O3 & Ethereum & General & DSL\\
Formality & Creates and analyses SCs with formal proofs of mathematical invariants about their behaviours.  & O3 & Ethereum & General & DSL\\
Functional solidity & Adds functional programming features to Solidity. & O5 & Ethereum & General & DSL\\
IELE & Formally verifies SCs and creates specifications that SCs must obey.  & O3 & Cardano & General & DSL\\
Ivy & Creates SCs for Bitcoin protocol and generating SegWit-compatible Bitcoin addresses. & O1 & Bitcoin , chain & General & DSL\\
L4  & Creates SCs for laws, regulations, business processes and business logic.   & O6 & Ethereum & Legal & DSL\\
Liquidity & Creates of SCs with security restrictions. & O2 & Tezos & General & DSL\\
LLL & Creates clean code. Enables developers to access SCs’ memory and opcodes. & O4 & Ethereum & General & DSL\\
Logikon & Enhances simplicity and security for SC implementation. & O2, O1 & Ethereum & General & DSL\\
Lolisa & Translates Solidity SCs into the formal language Lolisa and systematically extends them to other languages. & O3, O5 & Ethereum & General & DSL\\
Michelson & Creates SCs and provides proofs of their using formal verification. & O3 &  Tezos & General & DSL\\

\noalign{\smallskip}\hline
\end{tabular}
\end{table}

\begin{table}

\label{tab:17}       
\begin{tabular}{p{1.2 cm}p{4 cm}p{1cm }p{1cm }p{1.3 cm }p{1cm }}
\hline\noalign{\smallskip}
SCL Name & Pragmatics  & Objective & Target Platform & Application Domain & Domain specificity \\
\noalign{\smallskip}\hline\noalign{\smallskip}
Obsidian & Creates user-centered SCs with strong safety guarantees. & O1, O3 & Ethereum & General & DSL\\
Plutus  & Creates SCs addressing problems found in previous SCL designs.  & O3 & Plutus & General & DSL\\
Pyramid & Creates SCs and produces new SCLs that interoperate with every other SCL. & O5 & Ethereum & General & DSL\\

QSCL  & Creates a new kind of smart contract called Master Contract that can execute a contract triggered by off-chain and on-chain factors. & O1 & Qtum & General & DSL\\
Rholang  & Implements a contract language that generates semantics naturally suited for blockchain and SCs. & O3 & RChain & General & DSL\\
RIDE & Generates expression-based contracts with the ability to use external data in their logic if needed. & O1 & Waves & General & DSL\\
Scilla & Creates SCs with safety in mind.  & O1 & Zilliqa & General & DSL\\
Serpent & Creates SCs and blockchain code using low-level opcode manipulation; provides access to high-level primitives. & O4 & Ethereum  & General & DSL\\
Simplicity & Creates SCs for cryptocurrencies and blockchain applications. & O1 & Bitcoin , Ethereum  & General & DSL\\
Simvolio & Implements SCs with a minimum required number of program control commands and predefined functions. & O5 & Apla & General & DSL\\
Solidity & Generates executable smart contracts (SCs) and supports implementation. & O1  & Ethereum & General & DSL\\
SolidityX & Implements Secure-by-default SCs with performance optimization.  & O2, O4 & Ethereum & General & DSL\\
Viper/Vyper  & Implements human-readable SCs.  & O1 & Ethereum & General & DSL\\
Yul & Provides inline assembly for SCLs with different back-ends. & O3 & Ethereum & General & DSL\\

\noalign{\smallskip}\hline
\end{tabular}
\end{table}

\subsection{Analyzing the Scope of the Smart Contract Languages}
To answer RQ2, what defines the scope and goals of each of the surveyed SCLs, Table 6 shows a comparison between SCLs in terms of their pragmatics, objectives, application domains, domain specifications, and target platforms.

\subsubsection{Pragmatics}

Most SCLs focus on generating and implementing source code for smart contracts. Solidity and Michelson generate executable smart contracts that can be deployed directly to the blockchain. Similarly, Viper is designed to generate executable smart contracts but with maximal human-readable code. Bitcoin Script is used to write smart contracts (i.e., conditions) to control transferring Bitcoins. Logikon and eWASM emphasize code readability and simplicity, so that users from various backgrounds can access and understand a smart contract’s source code.

Only a few SCLs use graphical representation to reduce ambiguity and enhance the usability of smart contracts. For example, Babbage proposed visual “mechanical parts” (pipes, valves, rods, and levers) to represent smart contract components. These parts can interact to perform contracting tasks. This helps users (i.e., participants) who have limited knowledge of programming languages, but who need to understand the contracts they are involved in. 
Bamboo focuses on clearly specifying the behavior of smart contracts by forcing developers to specify the application using state machines and making the state transition explicit. This allows developers to be aware of the states of the implemented smart contract. The developer also can determine external calls from the contract by following the smart contract states that facilitate controlling the smart contract. Pyramid patterns also support the separation of state-changing and static functions in smart contracts, as Bamboo does~\cite{harz}.

\begin{figure*}

  \includegraphics[width=0.95\textwidth]{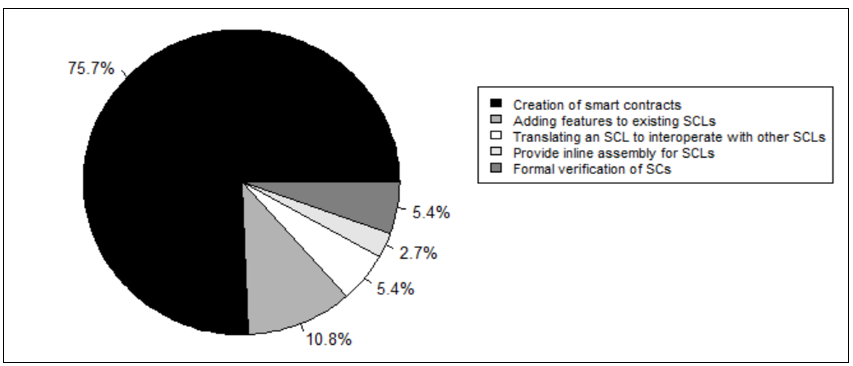}

\caption{An outline of surveyed SCLs’ pragmatics.}
\label{fig:5}       
\end{figure*}
\begin{figure*}

  \includegraphics[width=0.95\textwidth]{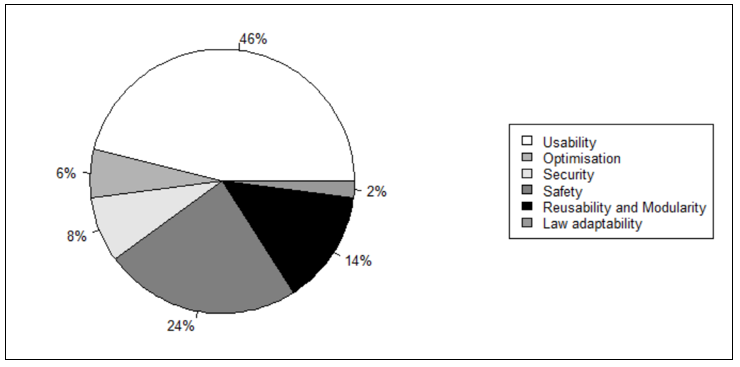}

\caption{An overview of surveyed SCLs’ objective.}
\label{fig:6}       
\end{figure*}

RIDE is another SCL that was implemented to be simple and accessible for novice and non-technical users who are not familiar with a particular language paradigm. RIDE prevents the usage of complex code blocks as smart contracts’ complexity cannot be predicted in advance, in this way this makes it possible to the developers to implement the smart contracts in a finite number of steps~\cite{Waves}.
Similar to RIDE, Obsidian offers a user-centered approach to designing smart contacts, to make it easier for programmers to write correct programs, while leveraging a type system to provide strong safety guarantees. Obsidian also serves as a testbed for research on language design methodology. However, Obsidian suggests that it is better to take users into account directly. Hence, Obsidian mainly uses formative user studies to inform language design changes that improve usability. 

Simplicity was intended as an advancement over existing SCLs, avoiding their shortcomings and opening new possibilities for smart contracts on blockchains, while allowing programmers to better verify safety and security. The Scilla, Flint, and Plutus contract languages impose structures on smart contracts that make applications less vulnerable to attack, by eliminating certain known vulnerabilities directly at the language level (e.g., re-entrancy and unhandled exceptions). Liquidity is a high-level statically typed language to program smart contracts. It uses a subset of the syntax of OCaml, and strictly complies to Michelson security restrictions. The E language implements smart contracts based on a predefined object model of secure, distributed, persistent computation, to guarantee intra-process security and the ability to run untrusted code safely.

In addition to improving safety and security, several languages aim to improve the anatomy and structural elements of the smart contract. ADICO explores and maps the structural elements and grammar of contracts, to simplify their creation. Rholang attempts to provide proper semantics. Simvolio reduces the number of commands and predefined functions to a minimal functional set. QSCL is structured around a Master Contract that executes contracts triggered by off-chain and on-chain factors. 

Some SCLs aim to create smart contracts that compile and extend the properties of other SCLs that can be deployed directly to the blockchain. For instance, Fi and Liquidity both support the creation of smart contracts that compile directly to valid and well-typed Michelson. Yul (previously called JULIA or IULIA) also compiles to several back-ends such as Solidity and eWASM. Its design allows it to be used for inline assembly in any future versions of Solidity. SolidityX and functional Solidity languages are designed to extend Solidity features such as language security and self-descriptive features. Lolisa extends Solidity to other languages by translating contracts from Solidity to formal Lolisa.  Lisp Like Language (LLL) facilitates the creation of very clean Solidity and Viper code, whilst removing the worst of the pains of coding such as stack management and jump management. Serpent implementations have produced insecure smart contracts, so Solidity, Viper or LLL are preferred instead~\cite{lll2}.

Languages such as Formality and IELE employ mathematical standards to create, analyze, and verify contracts formally. Formality uses formal proofs to create contracts and mathematical invariants to prove the integrity of their behavior analysis.
There are languages optimized for smart contracts used in regulatory and legal matters, financial derivatives, and business process management. These include L4, FirmoLang, and Findel. L4 attempts to capture the particularities of law, FirmoLang verifies financial contracts via a diminishing margin of error, and Findel models financial derivatives. The main pragmatic goals for all the languages mentioned above are broadly summarized in Figure 5.

\subsubsection{Objectives}

SCL designers describe many objectives that motivate their efforts. Most developers of the languages reviewed in our study (24 of 36) declared that better usability, for both developers and end-users, was their language design objective. There were 7 SCLs that described safety as an objective and 7 SCLs emphasized improved reusability and modularity. Only 4 SCLs focused on security during implementation as an objective. One single SCL emphasized adaptability (see Figure 6).

It is troubling to see that so few SCL designers have focused on smart contract security during implementation. Given the sensitive nature of many smart contract implementations, security should be a higher priority in the SCL design field. 
We also found that few languages provide ways of optimizing the implementation of smart contracts. This, again, is considered an unfortunate weakness of existing SCLs.

\begin{figure*}

  \includegraphics[width=0.75\textwidth]{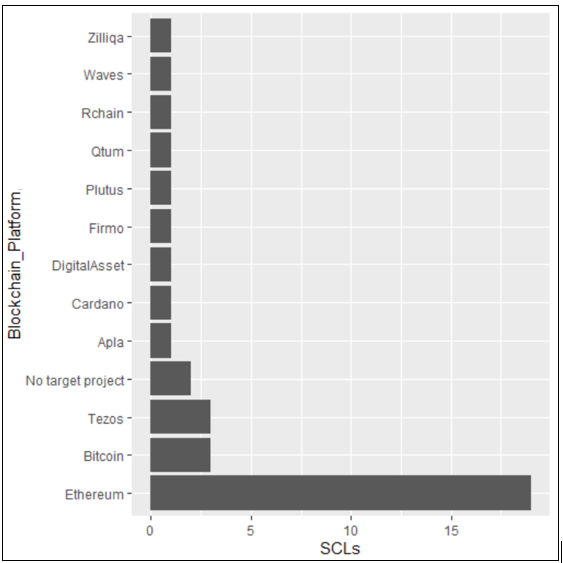}

\caption{An outline of surveyed SCLs’ target platforms.}
\label{fig:7}       
\end{figure*}

\subsubsection{	Target Platform}
Ethereum was the target platform for just over half (54\%) of the SCLs reviewed in this study. The SCLs are used to create smart contracts that get compiled into bytecode and executed on Ethereum Virtual Machine (EVM). Another class of SCLs are those that were originally designed for a specific platform, but they came to be used across other platforms. For example, Solidity is used on several private blockchain platforms including Monix \footnote{https://monix.io}, and Counterparty~\cite{counterparty}. As a platform, Bitcoin is a popular target for several SCLs, such as Bitcoin Script, Simplicity, Ivy, and BitML. Tezos is another popular platform that is a target for three SCLs (i.e., Michelson, Fi, and Liquidity). The remaining set (just under half) of the SCLs we reviewed were integrated with the blockchain platform they were designed for. For example, Simvolio is used to develop smart contracts on the Apla blockchain platform. RIDE targets waves’ platform, and the Plutus contract language is specifically for the Plutus Blockchain. Figure 7 shows the distribution of SCLs based on the target blockchain platform.

\subsubsection{	Application Domain}
SCLs can be used to implement either general or specific types of smart contracts. Most SCLs are general and can be used to implement smart contracts in any domain. Notable exceptions include DAML and Findel, which are specifically for financial contracts and are influenced by the financial products markup language \footnote{https://www.fpml.org/} as shown in Table 6 part (a) and part (b). Similarly, L4 language is for smart legal contracts that use legal directives. L4 is still in its early stages with no real-world applications yet.

\subsubsection{	Domain specificity}
We defined two domains for classifying an SCL. It can either be a General-Purpose Language (GPL) or a Domain Specific Language (DSL). DSLs are specifically designed for programming smart contracts alone. GPLs are languages that can be used to create several types of programs, including smart contracts.  

\subsection{Language Characteristics}
The smart contract language characteristics focus on the features and properties of the language. In this survey, we explore the syntax, semantics, level of abstraction, and programming paradigm, in addition to the language support of type checking and whether it is Turing complete or not.

\subsubsection{Syntax}
The abstract syntax of SCLs represents the concepts (or constructs) underlying the language and how these concepts/constructs are linked and related to each other. Every SCL has a concrete syntax and an abstract syntax. Both are essential parts of the language definition. In this work, we reviewed both the concrete and abstract syntax of all 36 languages. We divided concrete syntax into two categories: text-based and graphical. We found that concrete syntax often mirrored the constructs of popular GPLs, so we included that in our comparison. For abstract syntax, we examined three main abstract syntax attributes in our classification scheme related to programming and structuring smart contracts; those are design patterns, loop and recursion support, and language-specific constructs.

\begin{itemize}
    \item	\textbf{Design patterns}\\
   Design patterns describe elegant solutions to specific problems. An SCL either can have predefined code patterns that developers can use without the need to implement them from scratch, or it does not provide predefined code patterns. Predefined code patterns are reusable code idioms (e.g., functions, classes) that are implemented and ready-to-use solutions. For example, a language might support security code patterns (e.g., control access code pattern) and provide predefined idioms to address some of the common security pitfalls.
    \item\textbf{	Loop and Recursion support}

Our classification scheme singled out loop and recursion programming structures as particularly important for SCL classification, since they have been exploited multiple times in smart contracts attacks like reentrancy attacks~\cite{atzei2017survey}. In our classification, an SCL can either support looping and recursion or not.  
     \item	\textbf{Language specific constructs}\\
An SCL provides developers with a specific set of programming components that they can use in order to implement smart contract functionality. These programming components differ from one SCL to another. In this paper, we summarize all the programming components each SCL allows. 
    
\end{itemize}

\subsubsection{Semantics}
The semantics of any programming language give meaning to syntactically correct components and expressions~\cite{briscoe2011introduction}. SCL semantics generally involves transforming the correct syntactic components into a determined semantic domain. In this work, we classified semantics as either formal or informal. Formal semantics defines the program's behavior in a well-formed, unambiguous manner; unfortunately, they are typically harder to use and understand by the novice programmer. On the other hand, in informal semantics, the program requires interpretation, where compilers define the program's behavior. 

\subsubsection{Language level}
SCLs can be classified into three different levels: high-level languages, Intermediate Representation (IR) languages, and low-level languages. A high-level language is close to human language and is easily understood. A low-level language is a language with no abstraction, and its functions are very close to the processor instructions.  An IR language is intermediate between high-level and low-level languages and is mostly used to implement optimized programs and/or reasoning about source code properties such as code safety. 

\subsubsection{Type checking}
Checking and enforcing type constraints is known as type checking. It can be classified as static or dynamic. Static type checking is performed at compile-time. Dynamic type checking takes place at run-time. Some languages combine them both.

\subsubsection{Turing completeness}
One profound decision in the field of blockchain design is the property of Turing completeness. A programming language is Turing complete if it has the ability to simulate any single-tape Turing machine. In other words, it has a finite set of states, a defined way to change state, and reusable storage for reading and writing operations. We classify SCLs as either Turing complete or Turing incomplete. Moreover, we do not consider the role of the execution time environment in limiting or expanding the functionality of the language.    
\subsubsection{Programming paradigm}
In the course of our analysis, we observed that SCL designers describe programming paradigms in their documentation. We collected these descriptions and classified them as follows.
\begin{itemize}
 \item \textbf{Contract-oriented: } Languages that focus on client-supplier relationships between software modules by formally and precisely specifying the interfaces between those models.
 \item	\textbf{Calculus-based:}  Formal languages that use logical expressions and mathematical proofs to express computations and business logic. 
 \item\textbf{Functional programming:} A declarative programming paradigm that expresses the program logic and computations as pure functions and avoids mutable data. 
 \item\textbf{Imperative programming:} The program is expressed as a sequence of statements that specify how to reach specific goals. An algorithmic sequence of commands and instructions is needed to update the contract's state in explicit steps.
 \item\textbf{Secure by design:} Languages designed with significant effort applied to provide constructs that enable programmers to write programs that are free of vulnerabilities and bugs.  
 \item\textbf{Object-oriented programming:} Languages that decompose applications into objects that encapsulate data fields with the procedures that work on these data fields.
 \item	\textbf{Stack-oriented:} Languages based on a stack machine model for passing parameters. A stack-based language can have more than one stack, and each stack might serve a different purpose.  

\end{itemize}

\begin{table}

\caption{A summary of available SCLs.}
\label{tab:18}       
\begin{tabular}{p{2 cm}p{2.7 cm}p{2cm }p{3cm }}
\hline\noalign{\smallskip}
SCL Name & References  & Project Type & Current State  \\
\noalign{\smallskip}\hline\noalign{\smallskip}
ADICO &~\cite{adico} & Academic & Prototype\\
Babbage &~\cite{babbage}  & Not specified & Prototype\\
Bamboo &~\cite{Pirapira}  & Industrial & Alpha state\\
Bitcoin Script &~\cite{Script} & Industrial & Stable\\
BitML &~\cite{atzei2019developing} & Academic & Under development\\
DAML &~\cite{daml} & Industrial & Stable \\
E &~\cite{ERights} & Academic & Stable\\
eWASM &~\cite{ewasm} & Industrial & Under development\\
Fi &~\cite{goodman2014tezos} & Industrial & Under development\\
Findel &~\cite{biryukov2017findel} & Academic & Under development\\
FirmoLang &~\cite{FirmoLang} & Industrial & Under development\\
Flint &~\cite{schrans2018writing} & Industrial & Alpha state\\
Formality &~\cite{moonad} & Industrial & Under development\\
Functional solidity &~\cite{Raineorshine} & Not specified & Prototype\\
IELE &~\cite{kasampalis2019iele} & Academic & Prototype\\
Ivy &~\cite{Ivy-Lang} & Academic & Prototype\\
L4  &~\cite{Legalese} & Academic & Alpha state\\
Liquidity &~\cite{OCamlPro} & Industrial & Under development\\
LLL &~\cite{LLL} & Industrial & Under development\\
Logikon &~\cite{Logikon} & Industrial & Experimental\\
Lolisa &~\cite{yang2018lolisa} & Academic & Under development\\
Michelson &~\cite{goodmanimichelson} & Industrial & Under development\\
Obsidian &~\cite{coblenz2017obsidian}& Academic & Stable\\
Plutus   &~\cite{Plutus} & Both & Under development\\
Pyramid &~\cite{Pyramid} & Industrial & Stable\\
QSCL  &~\cite{Qtum} & Industrial & Stable\\
Rholang  &~\cite{eykholt2017rchain} & Industrial & Under development\\
RIDE &~\cite{Waves}  & Both & Under development\\
Scilla &~\cite{sergey2018scilla} & Academic & Under development\\
Serpent &~\cite{delmolino2015programmer} & Industrial & Stable\\
Simplicity &~\cite{o2017simplicity} & Academic & Under development\\
Simvolio &~\cite{Simvolio} & Industrial & Prototype\\
Solidity &~\cite{solidity,modi2018solidity,Solidity2} & Industrial  & Stable\\
SolidityX &~\cite{Solidityx} & Industrial & Beta state\\
Viper/Vyper &~\cite{Viper} & Industrial & Beta state\\
Yul &~\cite{Yul} & Industrial & Under development\\

\noalign{\smallskip}\hline
\end{tabular}
\end{table}
\section{Survey Results and Discussion}
In this section, we present a full comparison between the 36 SCLs we investigated, using our proposed framework. This section is organized to address the research questions (RQs) we introduced in Section 3.1.
\subsection{The Smart Contract Languages Surveyed}

\begin{figure*}

  \includegraphics[width=0.95\textwidth]{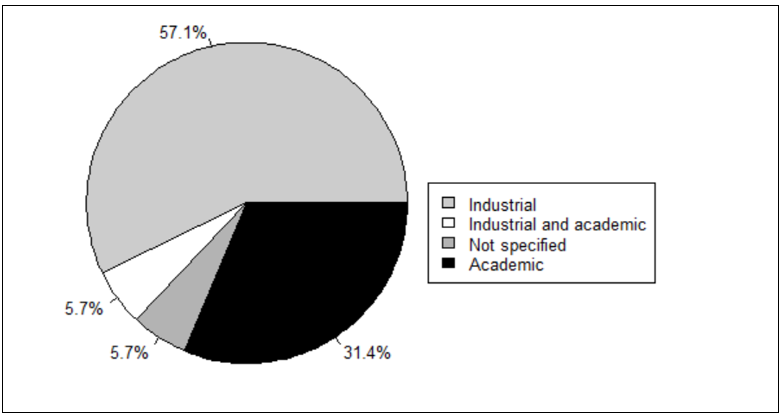}

\caption{The distribution of surveyed SCLs based on whether they are created as part of academic or industry based projects.}
\label{fig:8}       
\end{figure*}
\begin{figure*}

  \includegraphics[width=0.95\textwidth]{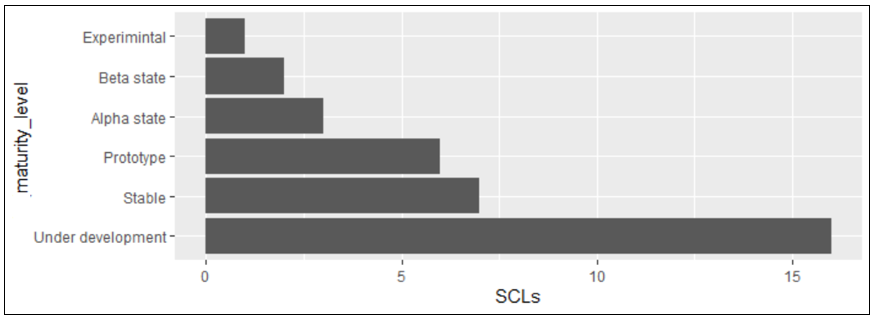}

\caption{The maturity levels of the surveyed SCLs.}
\label{fig:9}       
\end{figure*}

To answer the question of what are the common existing SCLs (RQ1), we have tabulated the list of the 36 SCL we surveyed in Table 5, classified by type of project (academic, industry), and the maturity level of the project (i.e., its current development state). As shown in Figure 3, more than half of these surveyed languages are industry-based projects. Eleven out of 36 SCLs (i.e., around 31\%) were created or defined as academic research projects. Only about 6\% of the surveyed languages were collaborations between industry and academia.

We also classified the surveyed SCLs based on their maturity level and current state. A stable language is a language that is being actively used to create smart contracts, and that is at high maturity levels. We found 8 stable SCLs: Solidity, Obsidian, Pyramid, Digital Asset Modeling Language (DAML), Bitcoin Script, E, Qtum Smart Contract Language QSCL, and Serpent. Alpha state languages included Bamboo, L4, and Flint. Alpha is the first stage of testing a language, and it is the first phase in the language development life cycle. Viper and SolidityX are in the beta stage, which is the second phase in the development of a programming language. Six languages are prototype languages: Functional solidity, Babbage, Intermediate-Level Blockchain Language (IELE), Ivy, Simvolio, and ADICO. A prototype language provides a model that allows developers to work on object cloning and prototyping. Logikon is an experimental language. The rest of SCLs are still under development (see Figure 4).

To sum up, we reviewed the available SCLs in this section, their target fields (industry or academy), and their maturity levels. We found that most of these SCLs target the industrial field. As well as most of these SCLs are still in the development phase. Only 7 out of 36 SCLs are stable by the time of writing this survey.  

\begin{table}

\caption{A comparison of the scope of SCLs.}
\label{tab:19}       
\begin{tabular}{p{1.2 cm}p{4 cm}p{1cm }p{1cm }p{1.2 cm }p{1cm }}
\hline\noalign{\smallskip}
SCL Name & Pragmatics  & Objective & Target Platform & Application Domain & Domain specificity \\
\noalign{\smallskip}\hline\noalign{\smallskip}
ADICO & Creates mapping methods that simplify the generation of Smart Contracts (SCs). & O1 & No project & General & DSL\\
Babbage & Creates SCs using a visual mechanical metaphor. & O1 & Ethereum & General & DSL\\
Bamboo & Implements SCs improving reentrancy behavior and formal verification.  & O2 & Ethereum & General & DSL\\
Bitcoin Script& Allow users to specify certain conditions (contracts) that determine how spending Bitcoins shall be done and who shall have access to the Bitcoins being transferred in the network& O1 & Bitcoin & General & DSL \\
BitML & Creates SCs that regulate transfer of bitcoins among participants. & O1 & Bitcoin & General & DSL\\
DAML & Implements distributed ledgers and smart contracts, enables high-level specification of real-time business logic. & O3  & Digital Asset  & Financial & DSL\\
E & Object modelling of secure SCs and distributed persistent computation. & O5 & No project  & General & DSL\\
eWASM & Generates SCs with accessible development for users. & O1 & Ethereum  & General & DSL\\
Fi & Creates SCs that compile directly to valid and well-typed Michelson code. & O5 & Tezos & General & DSL\\
Findel & Models financial derivatives on blockchains and SCs. & O1 & Ethereum  & Financial & DSL\\
FirmoLang & Creates verified financial contracts with a diminishing margin of error in automated SCs & O3 & FIRMO & General & DSL\\
Flint & Creates robust SCs with contract safety in mind. & O3 & Ethereum & General & DSL\\
Formality & Creates and analyses SCs with formal proofs of mathematical invariants about their behaviours.  & O3 & Ethereum & General & DSL\\
Functional solidity & Adds functional programming features to Solidity. & O5 & Ethereum & General & DSL\\
IELE & Formally verifies SCs and creates specifications that SCs must obey.  & O3 & Cardano & General & DSL\\
Ivy & Creates SCs for Bitcoin protocol and generating SegWit-compatible Bitcoin addresses. & O1 & Bitcoin , chain & General & DSL\\
L4  & Creates SCs for laws, regulations, business processes and business logic.   & O6 & Ethereum & Legal & DSL\\
Liquidity & Creates of SCs with security restrictions. & O2 & Tezos & General & DSL\\
LLL & Creates clean code. Enables developers to access SCs’ memory and opcodes. & O4 & Ethereum & General & DSL\\
Logikon & Enhances simplicity and security for SC implementation. & O2, O1 & Ethereum & General & DSL\\
Lolisa & Translates Solidity SCs into the formal language Lolisa and systematically extends them to other languages. & O3, O5 & Ethereum & General & DSL\\
Michelson & Creates SCs and provides proofs of their using formal verification. & O3 &  Tezos & General & DSL\\

\noalign{\smallskip}\hline
\end{tabular}
\end{table}

\begin{table}

\label{tab:2}       
\begin{tabular}{p{1.2 cm}p{4 cm}p{1cm }p{1cm }p{1.3 cm }p{1cm }}
\hline\noalign{\smallskip}
SCL Name & Pragmatics  & Objective & Target Platform & Application Domain & Domain specificity \\
\noalign{\smallskip}\hline\noalign{\smallskip}
Obsidian & Creates user-centered SCs with strong safety guarantees. & O1, O3 & Ethereum & General & DSL\\
Plutus  & Creates SCs addressing problems found in previous SCL designs.  & O3 & Plutus & General & DSL\\
Pyramid & Creates SCs and produces new SCLs that interoperate with every other SCL. & O5 & Ethereum & General & DSL\\

QSCL  & Creates a new kind of smart contract called Master Contract that can execute a contract triggered by off-chain and on-chain factors. & O1 & Qtum & General & DSL\\
Rholang  & Implements a contract language that generates semantics naturally suited for blockchain and SCs. & O3 & RChain & General & DSL\\
RIDE & Generates expression-based contracts with the ability to use external data in their logic if needed. & O1 & Waves & General & DSL\\
Scilla & Creates SCs with safety in mind.  & O1 & Zilliqa & General & DSL\\
Serpent & Creates SCs and blockchain code using low-level opcode manipulation; provides access to high-level primitives. & O4 & Ethereum  & General & DSL\\
Simplicity & Creates SCs for cryptocurrencies and blockchain applications. & O1 & Bitcoin , Ethereum  & General & DSL\\
Simvolio & Implements SCs with a minimum required number of program control commands and predefined functions. & O5 & Apla & General & DSL\\
Solidity & Generates executable smart contracts (SCs) and supports implementation. & O1  & Ethereum & General & DSL\\
SolidityX & Implements Secure-by-default SCs with performance optimization.  & O2, O4 & Ethereum & General & DSL\\
Viper/Vyper  & Implements human-readable SCs.  & O1 & Ethereum & General & DSL\\
Yul & Provides inline assembly for SCLs with different back-ends. & O3 & Ethereum & General & DSL\\

\noalign{\smallskip}\hline
\end{tabular}
\end{table}

\subsection{Analyzing the Scope of the Smart Contract Languages}
To answer RQ2, what defines the scope and goals of each of the surveyed SCLs, Table 6 shows a comparison between SCLs in terms of their pragmatics, objectives, application domains, domain specifications, and target platforms.

\subsubsection{Pragmatics}

Most SCLs focus on generating and implementing source code for smart contracts. Solidity and Michelson generate executable smart contracts that can be deployed directly to the blockchain. Similarly, Viper is designed to generate executable smart contracts but with maximal human-readable code. Bitcoin Script is used to write smart contracts (i.e., conditions) to control transferring Bitcoins. Logikon and eWASM emphasize code readability and simplicity, so that users from various backgrounds can access and understand a smart contract’s source code.

Only a few SCLs use graphical representation to reduce ambiguity and enhance the usability of smart contracts. For example, Babbage proposed visual “mechanical parts” (pipes, valves, rods, and levers) to represent smart contract components. These parts can interact to perform contracting tasks. This helps users (i.e., participants) who have limited knowledge of programming languages, but who need to understand the contracts they are involved in. 
Bamboo focuses on clearly specifying the behavior of smart contracts by forcing developers to specify the application using state machines and making the state transition explicit. This allows developers to be aware of the states of the implemented smart contract. The developer also can determine external calls from the contract by following the smart contract states that facilitate controlling the smart contract. Pyramid patterns also support the separation of state-changing and static functions in smart contracts, as Bamboo does~\cite{harz}.

\begin{figure*}

  \includegraphics[width=0.95\textwidth]{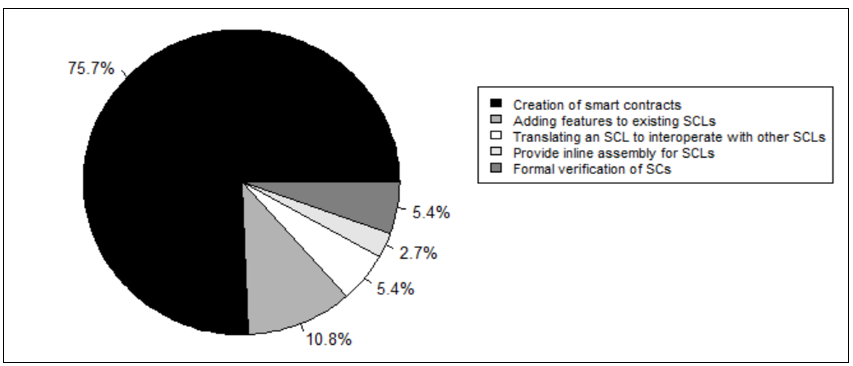}

\caption{An outline of surveyed SCLs’ pragmatics.}
\label{fig:12}       
\end{figure*}
\begin{figure*}

  \includegraphics[width=0.95\textwidth]{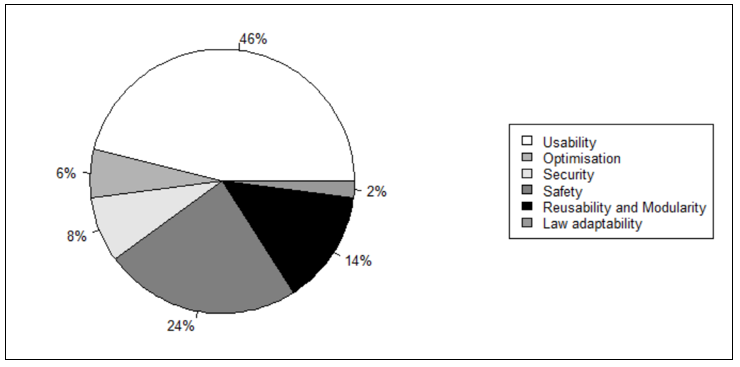}

\caption{An overview of surveyed SCLs’ objective.}
\label{fig:13}       
\end{figure*}

RIDE is another SCL that was implemented to be simple and accessible for novice and non-technical users who are not familiar with a particular language paradigm. RIDE prevents the usage of complex code blocks as smart contracts’ complexity cannot be predicted in advance, in this way this makes it possible to the developers to implement the smart contracts in a finite number of steps~\cite{Waves}.
Similar to RIDE, Obsidian offers a user-centered approach to designing smart contacts, to make it easier for programmers to write correct programs, while leveraging a type system to provide strong safety guarantees. Obsidian also serves as a testbed for research on language design methodology. However, Obsidian suggests that it is better to take users into account directly. Hence, Obsidian mainly uses formative user studies to inform language design changes that improve usability. 

Simplicity was intended as an advancement over existing SCLs, avoiding their shortcomings and opening new possibilities for smart contracts on blockchains, while allowing programmers to better verify safety and security. The Scilla, Flint, and Plutus contract languages impose structures on smart contracts that make applications less vulnerable to attack, by eliminating certain known vulnerabilities directly at the language level (e.g., re-entrancy and unhandled exceptions). Liquidity is a high-level statically typed language to program smart contracts. It uses a subset of the syntax of OCaml, and strictly complies to Michelson security restrictions. The E language implements smart contracts based on a predefined object model of secure, distributed, persistent computation, to guarantee intra-process security and the ability to run untrusted code safely.

In addition to improving safety and security, several languages aim to improve the anatomy and structural elements of the smart contract. ADICO explores and maps the structural elements and grammar of contracts, to simplify their creation. Rholang attempts to provide proper semantics. Simvolio reduces the number of commands and predefined functions to a minimal functional set. QSCL is structured around a Master Contract that executes contracts triggered by off-chain and on-chain factors. 

Some SCLs aim to create smart contracts that compile and extend the properties of other SCLs that can be deployed directly to the blockchain. For instance, Fi and Liquidity both support the creation of smart contracts that compile directly to valid and well-typed Michelson. Yul (previously called JULIA or IULIA) also compiles to several back-ends such as Solidity and eWASM. Its design allows it to be used for inline assembly in any future versions of Solidity. SolidityX and functional Solidity languages are designed to extend Solidity features such as language security and self-descriptive features. Lolisa extends Solidity to other languages by translating contracts from Solidity to formal Lolisa.  Lisp Like Language (LLL) facilitates the creation of very clean Solidity and Viper code, whilst removing the worst of the pains of coding such as stack management and jump management. Serpent implementations have produced insecure smart contracts, so Solidity, Viper or LLL are preferred instead~\cite{lll2}.

Languages such as Formality and IELE employ mathematical standards to create, analyze, and verify contracts formally. Formality uses formal proofs to create contracts and mathematical invariants to prove the integrity of their behavior analysis.
There are languages optimized for smart contracts used in regulatory and legal matters, financial derivatives, and business process management. These include L4, FirmoLang, and Findel. L4 attempts to capture the particularities of law, FirmoLang verifies financial contracts via a diminishing margin of error, and Findel models financial derivatives. The main pragmatic goals for all the languages mentioned above are broadly summarized in Figure 5.

\subsubsection{Objectives}

SCL designers describe many objectives that motivate their efforts. Most developers of the languages reviewed in our study (24 of 36) declared that better usability, for both developers and end-users, was their language design objective. There were 7 SCLs that described safety as an objective and 7 SCLs emphasized improved reusability and modularity. Only 4 SCLs focused on security during implementation as an objective. One single SCL emphasized adaptability (see Figure 6).

It is troubling to see that so few SCL designers have focused on smart contract security during implementation. Given the sensitive nature of many smart contract implementations, security should be a higher priority in the SCL design field. 
We also found that few languages provide ways of optimizing the implementation of smart contracts. This, again, is considered an unfortunate weakness of existing SCLs.

\begin{figure*}

  \includegraphics[width=0.75\textwidth]{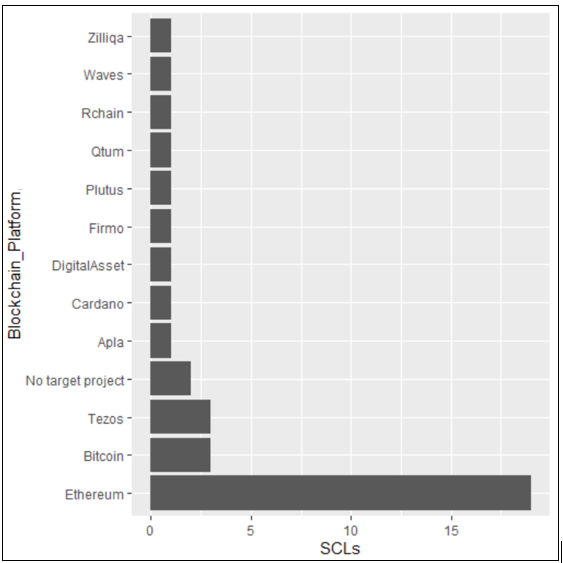}

\caption{An outline of surveyed SCLs’ target platforms.}
\label{fig:14}       
\end{figure*}

\subsubsection{	Target Platform}
Ethereum was the target platform for just over half (54\%) of the SCLs reviewed in this study. The SCLs are used to create smart contracts that get compiled into bytecode and executed on Ethereum Virtual Machine (EVM). Another class of SCLs are those that were originally designed for a specific platform, but they came to be used across other platforms. For example, Solidity is used on several private blockchain platforms including Monix \footnote{https://monix.io}, and Counterparty~\cite{counterparty}. As a platform, Bitcoin is a popular target for several SCLs, such as Bitcoin Script, Simplicity, Ivy, and BitML. Tezos is another popular platform that is a target for three SCLs (i.e., Michelson, Fi, and Liquidity). The remaining set (just under half) of the SCLs we reviewed were integrated with the blockchain platform they were designed for. For example, Simvolio is used to develop smart contracts on the Apla blockchain platform. RIDE targets waves’ platform, and the Plutus contract language is specifically for the Plutus Blockchain. Figure 7 shows the distribution of SCLs based on the target blockchain platform.

\subsubsection{	Application Domain}
SCLs can be used to implement either general or specific types of smart contracts. Most SCLs are general and can be used to implement smart contracts in any domain. Notable exceptions include DAML and Findel, which are specifically for financial contracts and are influenced by the financial products markup language \footnote{https://www.fpml.org/} as shown in Table 6 part (a) and part (b). Similarly, L4 language is for smart legal contracts that use legal directives. L4 is still in its early stages with no real-world applications yet.

\subsubsection{	Domain specificity}
Most existing SCLs are Domain Specific languages (DSL) designed specifically for smart contract implementation. However, some General Purpose Languages (GPL) are used to implement smart contracts. These include Idris \footnote{https://www.idris-lang.org/}, and Golang \footnote{https://golang.org/}. Some blockchain platforms use GPLs due to their popularity. For example, Hyperledger Fabric \footnote{https://www.hyperledger.org/projects/fabric} uses Golang for a vast portion of its smart contracts’ chaincode. However, among the 36 languages we reviewed, Formality and Rholang were the only GPLs used. When GPLs are used, smart contract asset and unit types have to be designed, and ledger states and SC libraries need to be imported through specific functions. Caution is needed when designing smart contracts in Formality and Rholang because they support unbounded loops and recursion. Aside from these 2 languages, the remaining 33 languages in our review are DSLs, as shown in Table 6.  

To recap, we have discussed the SCLs’ scope in this section, the pragmatics, objectives, and target projects for each SCLs (RQ2). We noted that SCLs have similar underlying pragmatics, even when they claim to be geared towards different objectives and goals. Those claims prompt users to use specific SCLs to achieve different objectives, based on supposedly unique advantages of each language. Only a few languages support smart contract source code transformation into other popular SCLs, and code abstraction and visualization approaches are barely supported by the available SCLs. This makes it hard for non-technical users to understand smart contracts.

\begin{table}

\caption{A comparison of the characteristics of SCLs.}
\label{tab:1}       
\begin{tabular}{p{1 cm}p{1 cm}p{2.3cm }p{1cm }p{1.5 cm }p{2.3cm }}
\hline\noalign{\smallskip}
SCL Name & Concrete Syntax & Syntax Influenced by &   & Abstract Syntax &  \\

& & & DPs & LR support & Main programming components\\
\noalign{\smallskip}\hline\noalign{\smallskip}
ADICO & Text & Nested ADICO (nADICO) & Yes & No & Statements, Conditions, “Or else” \\ 
Babbage & Graphics & Mechanical-parts & No & No & Pipes, valves, rods, levers \\
Bamboo & Text & Erlang & No & No & Functions \\
Bitcoin Script& Text& Forth language & No& No& Constants,Flow controls, Locktime, Splice, Arithmetic, Crypto operations \\
BitML & Text & process-algebraic language & No & No & Symbolic model computational model \\ 
DAML & Text & Financial products Markup Language (FpML) & Yes & No & Set of DAML templates and patterns \\ 
E & Text & Java & No & No & Functions and variables \\ 
eWASM & Text & WebAssembly & No & Yes & Functions and variables \\ 
Fi & Text & ECMAScript/ JavaScript and Solidity & No & No & Storage Variables, Structs and Contract Entry Points \\ 
Findel & Text & Financial products Markup Language (FpML) & No & No & Pre-defend gateway smart contract, balance tokens,  predefined derivatives \\ 
FirmoLang & Text & FpML and xml & No & No & Derivative With contract units,  \\ 
Flint & Text & Swift, Solidity & Yes & Yes & Protected blocks with  functions,  integrated design patterns \\ 
Formality & Text & Haskell & No & Yes,  Limited   & Datatypes, lazy copying, simple proofs \\ 
Functional solidity & Text & JavaScript  & No & No & Methods  represent contract states \\ 
IELE & Text & LLVM & Yes & No & Deposit Handler function, pre-defind functions  variables \\ 
Ivy & Text &  C and JavaScript & No & No & Clauses and pre-defend functions \\ 
L4  & Text & Verilog & No & No & In development \\ 
Liquidity & Text & OCaml & No & Yes & Local values definitions, initializer, entry points. \\
LLL & Text & LISP & No & No & Memory Layout Constructor, Functions \\ 
Logikon & Text & Prolog & No & No & Logical constraints \\ 
Lolisa & Text & JavaScript  & No & Yes & Subset of solidity\\
Michelson & Text & Bitcoin Script & No & Yes & Series of instructions\\
Obsidian & Text & JavaScript  & No & Yes, Limited   & Methods are invoked as object’s, current state\\

\noalign{\smallskip}\hline
\end{tabular}
\end{table}

\begin{table}

\label{tab:3}       
\begin{tabular}{p{1 cm}p{1 cm}p{2.3cm }p{1cm }p{1.5 cm }p{2.3cm }}
\hline\noalign{\smallskip}
SCL Name & Concrete Syntax & Syntax Influenced by &   & Abstract Syntax &  \\

& & & DPs & LR support & Main programming components\\
\noalign{\smallskip}\hline\noalign{\smallskip}

Plutus  & Text & Haskell-like & No & No & Functions and variables\\
Pyramid & Text & Racket  & No & Yes & Functions, variables\\
QSCL & Text & Solidity  & No & No & Who model, where model, what model\\
Rholang & Text &  $\rho$ -calculus & No & Yes & Persistent state, code, RChain address(es)\\
RIDE & Text & F\# and Scala  & No & No & Accounts, Assets and functions\\
Scilla & Text & Python & No & No & Implicit balance, variables and interfaces\\
Serpent & Text & Python & Yes & Yes & Same as in solidity\\
Simplicity & Text & Bitcoin Script & No & No & Set of functions, expressions combinators.\\
Simvolio & Text &  C language & No & Yes & Data conditions actions.\\
Solidity & Text & JavaScript  & Yes & Yes, & Variables, Functions, Modifiers, Events, Struct, Enum Types\\
 &  &  &  & limited  & \\
SolidityX & Text & JavaScript  & No & Yes & Same as in solidity\\
Viper & Text & Python  & Yes & Yes ,   limited  & File contains declarations of State Variables, Functions, Structs\\
Yul & Text & JavaScript  & No & Yes & Functions, blocks, variables, literals, if and  Switch, expressions\\

\noalign{\smallskip}\hline
\end{tabular}
\begin{tablenotes}
\item LEGEND: DPs refers to the Design Patterns, LR support is the Loop \& recursion support in each language.  
\end{tablenotes}
\end{table}

\subsection{The Characteristics of Smart Contract Languages}
To analyze the characteristics of each of the SCLs from a language engineering perspective (RQ3), we compared the main characteristics of the available SCLs in Table 7 and Table 8. 

\subsubsection{Syntax}
In this subsection, we discuss the concrete syntax and abstract syntax of the SCLs surveyed.
\begin{itemize}

\item \textbf{Concrete Syntax:} Most of the SCLs we surveyed used textual notation for their concrete syntax (see Table 7)). Only one SCL, Babbage, used graphics instead of text to make smart contract implementation easier. The concrete syntax of many of the SCLs surveyed resembled JavaScript and benefited from its simplicity, interoperability, and ubiquity\/popularity~~\cite{brooks2007fundamentals}. Java, C, and Python also influenced the syntax of some SCLs that follow the object-oriented programming paradigm. The syntax of a few SCLs was influenced by other SCLs, such as Flint with swift; and Solidity and RIDE with F\# and Scala. Simplicity is influenced by Bitcoin Script.
    
    \item\textbf{Abstract Syntax}     
  We looked at three classes of the abstract syntax of smart contracts that each surveyed SCL provides to developers (again, see Table 7). We illustrate contract types supported by the SCLs to see if they provide predefined patterns for implementing their contracts. Then we discuss the language-specific constructs that help developers, and we examine these languages support looping and recursion or not.  

\begin{figure*}

  \includegraphics[width=0.95\textwidth]{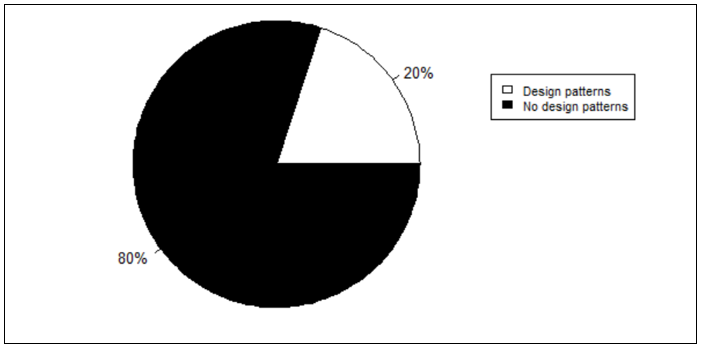}

\caption{The distribution of SCLs that provide design patterns among all the surveyed SCLs.}
\label{fig:15}       
\end{figure*}

     \item 	\textbf{Design Patterns} \\      
Figure 8 shows the distribution of the SCLs that provide developers with design patterns, relative to all reviewed SCLs. The figure shows that only seven SCLs (20\%) ship with design patterns. This is unfortunate because smart contracts need to utilize best practices to keep them correct and secure against attacks. 

\begin{figure*}

  \includegraphics[width=0.95\textwidth]{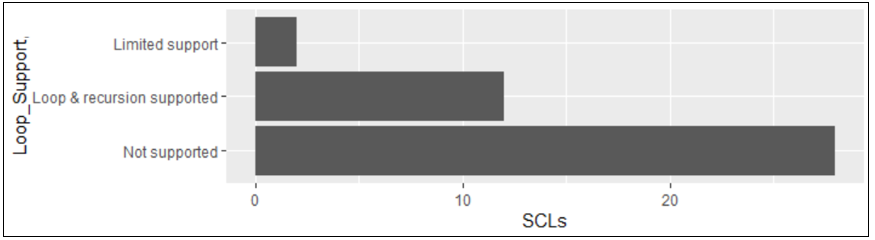}

\caption{Supporting loops and recursion in the surveyed SCLs.}
\label{fig:16}       
\end{figure*}

   \item \textbf{Loop\& recursion support}

Figure 9 shows that there are 12 SCLs that support loops and recursion with no restrictions. Two languages allow restricted use of loops and recursion; the rest exclude them outright. Unfortunately, while many successful attacks on smart contracts have taken advantage of unbounded recursion (e.g., DAO attack), most SCLs (i.e., 29 language) do not support loops and recursions due to the risk when using them without restriction in smart contracts.
   
   \item \textbf{	Language specific constructs}\\
 The language-specific constructs of a smart contract consist of functions, variables, and expressions. Additionally, some languages provide unique structural constructs to facilitate the implementation process. We will discuss the unique constructs in some languages. In this section, any SCLs we do not mention are similar in how they provide language constructs such as functions, variables, and expressions. We explicitly summarize all these language constructs in Table 7 for each SCL.

\begin{figure*}

  \includegraphics[width=0.95\textwidth]{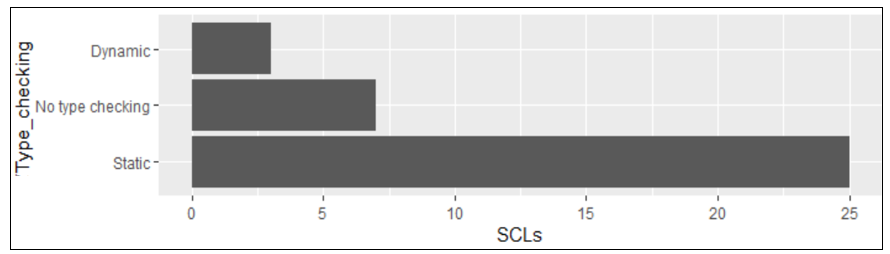}

\caption{A summary of SCLs’ type checking.}
\label{fig:17}       
\end{figure*}
One language that provides unique specific constructs to developers is DAML. It offers predefined templates that can be used to implement correct smart contracts. Only a few languages, such as IELE and Findel, provide predefined functions that the programmer can use directly to achieve certain programming tasks. Examples of these functions include predefined handler functions in IELE and predefined gateway smart contracts in Findel.  Using predefined functions to call external functions or other contracts is still critical due to unpredictable behavior and state changes.  A clear formal definition of these functions and their semantics would be beneficial to the programmers. In addition, it would elucidate state changes using explicit finite-state automata to reduce the risk of unexpected behavior during run time.
    
\end{itemize}
\begin{figure*}

  \includegraphics[width=0.95\textwidth]{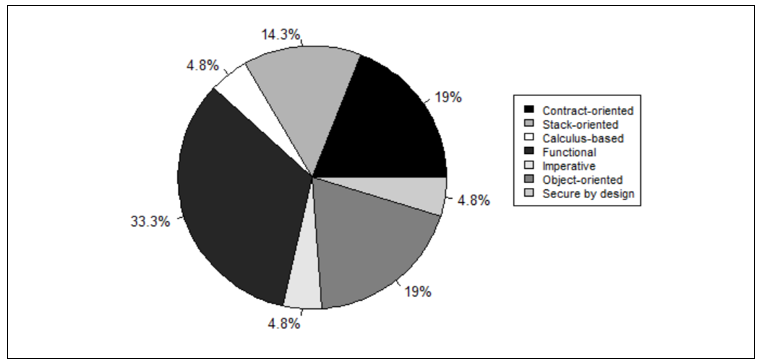}

\caption{An overview of other programming paradigms used in SCLs.}
\label{fig:18}       
\end{figure*}

\begin{table}

\caption{A comparison of the characteristics of SCLs.}
\label{tab:4}       
\begin{tabular}{p{1.2 cm}p{2 cm}p{2cm }p{1cm }p{1.3 cm }p{2cm }}
\hline\noalign{\smallskip}
SCLs name &Semantics  &Language-level  & Type checking  & Turing Completeness & Programming paradigm\\
\noalign{\smallskip}\hline\noalign{\smallskip}

ADICO & N/A & high-level  & Static & No & No other  paradigm \\
Babbage & In progress & high-level  & N/A & No & No other  paradigm \\
Bamboo & Informal & high-level  & N/A & Yes & Functional  \\
Bitcoin Script & informal & high-level& No & No & stack-oriented\\ 
BitML & Operational semantic  & high-level  & Static & Yes & Calculus -based  \\
DAML & N/A & high-level  & Static & No & Functional \\
E & Informal & high-level  & N/A & No & Object-oriented \\
eWASM & Informal & high-level  & N/A & No & Stack-oriented \\
Fi & Informal & high-level  & Static & No & Object-oriented \\
Findel & Informal & high-level  & Static & No & No other  paradigm \\
FirmoLang & Formal  & high-level  & Static & Yes & No other  paradigm \\
Flint & In progress & high-level  & Static & Yes & Contract-oriented  \\
Formality & Informal & high-level  & N/A & Yes & No other  paradigm \\
Functional solidity & Informal & high-level  & Static & Yes & No other  paradigm \\
IELE & Formal  & Intermediate & N/A & Yes & N/A\\
Ivy & Formal & high-level  & Static & No & Imperative \\
L4  & Informal & high-level  & Static & Yes & No other  paradigm \\
Liquidity & In progress & high-level  & Static & No & Functional \\
LLL & Informal & low-level & Static  & No & Stack-oriented \\
Logikon & Informal & high-level  & Static & Yes & Functional \\
Lolisa & Formal & high-level  & Static & Yes & No other  paradigm \\
Michelson & In progress & low-level  & Static & No & Stack-oriented \\
Obsidian & Informal & high-level  & Static & No & Object-oriented \\
Plutus language  & Informal & high-level  & Static & No & Contract-oriented \\
Pyramid & Informal & high-level  & Static & Yes & Functional \\
QSCL  & Formal & high-level  & N/A& Yes & No other  paradigm \\
Rholang  & In progress & high-level  & N/A& Yes & Functional \\
RIDE & Informal & high-level  & Static  & No & No other  paradigm \\
Scilla & Formal  & Intermediate & Static & No & No other  paradigm \\
Serpent & Informal & high-level  & Dynamic & Yes & No other  paradigm \\
Simplicity & Formal & low-level & Static & No & Functional \\
Simvolio & Informal & high-level  & Dynamic & Yes & No other  paradigm \\
Solidity & Informal & high-level  & Static  & Yes* & Contract-oriented \\
SolidityX & Informal & high-level  & Static & Yes & Secure-by-design \\
Viper & Informal & high-level  & Static & No & Contract-oriented \\
Yul & Informal & Intermediate  & Dynamic & Yes & Object-oriented \\

\noalign{\smallskip}\hline
\end{tabular}
\begin{tablenotes}
\item We refer to unavailable information with N/A.  
\end{tablenotes}

\end{table}

\subsubsection{Semantics}
As shown in Table 8, only a few SCLs (25\%) have formal semantics, including Simvolio, Pact, QSCL, Simplicity, Ivy, FirmoLang, Scilla, IELE, and Lolisa. Some SCLs are still in the process of formalizing their semantics and will officially release it as announced in their documented development timelines. Most SCLs use informal semantics, as shown in Table 8, so the smart contract's correct interpretation is left to the compiler. However, there have been attempts to formalize the semantics of some of these languages. For instance, Solidity has been formally defined in K~\cite{jiao2018executable}, Lem~\cite{hirai2017defining}, and F*~\cite{grishchenko2018semantic}.

\subsubsection{	Language-level}
More than a third of the reviewed SCLs are high-level languages that provide human-readable syntax to express a smart contract. Three out of the 36 reviewed SCLs are low-level languages (i.e., Michelson, Simplicity, and LLL). These languages implement deterministic contracts that are stored on the blockchain in the low-level format that will be executed by the distributed VM. For example, LLL is a low-level language similar to assembly language that implements low-level contracts and enables them to use the EVM’s limited resources efficiently. There are only three intermediate representation languages among the SCLs we reviewed: IELE, Scilla, and Yul. Intermediate representation enables reasoning about features such as the security and safety of contracts. A classification of all the reviewed languages based on language level is provided in Table 8.

\subsubsection{	Type checking}
Table 8 compares the type-checking mode - static or dynamic - of the SCLs we reviewed. Most SCLs use static type checkers (see Figure 10). Only three languages use dynamic type checking. To the best of our knowledge at the time of writing this survey, no SCL among those surveyed uses both static and dynamic type checkers. We also noticed that 7 languages have no type checking, probably because they are still under development.

\subsubsection{Programming paradigm}

We collected language paradigm information from SCL documentation, and we omitted the languages that did not declare their paradigm. Figure 11 shows that almost a third (33\%) of the surveyed SCLs have adopted a variation of a functional programming approach. Contract-oriented and object-oriented paradigms were represented in equal numbers (19\%). The stack-oriented paradigm was observed in 14\% of the languages. One language was imperative, one calculus-based, and one was secure-by-design (each representing 5\% of the total).

\subsubsection{Turing completeness}
Almost half of the surveyed SCLs (54\%) are Turing complete languages, while the rest are Turing incomplete. The almost equal split reflects the debate within the blockchain community on whether implementing smart contracts needs Turing complete languages or not~\cite{jansen2019smart}. Smart contracts written using Turing complete languages can specify arbitrary computations and execute complex functions. This gives programmers flexibility to implement the functionality they desire (e.g., it provides means for loops and/or complex recursive calls). However, when using Turing complete languages, smart contract specifications can only be validated using the fallible processes of code review and program inspection. As a result, the more complex a smart contract implementation is, the more error-prone it will be. Many failure incidents have occurred because of the complexity of the implementation of smart contracts and the inability to determine their semantics before execution (e.g., the DAO attack).

Turing incomplete languages is less complex, providing basic flow control statements like conditionals. This allows programmers and users to verify the execution flow before execution. However, language designers have already started to follow new approaches and find solutions to overcome the problems resulting from Turing complete languages (i.e., the pseudo-Turing complete approach~\cite{prieto2019blockchain}). These new solutions are mainly based on two ideas: (1) restricting loops and recursions for smart contracts using the run time environment (e.g., Ethereum)~\cite{prieto2019blockchain}), and (2) unwinding recursive calls between multiple transactions and blocks instead of using a single one. Even Solidity is a Turing complete language, but the run time environment of Ethereum, in which Solidity-based smart contracts are executed, limits loops and recursions for these smart contracts. One advantage of this approach is that developers have a great deal of flexibility in creating the functionality implemented in smart contracts.
Unlike Solidity, RIDE is a Turing incomplete language, and it does not support any means of loop or recursion. It provides a simple programming structure that can be tested and verified easily against security flaws. It also allows developers to estimate the computational effort needed to execute the smart contract. Still, RIDE can become Turing complete when it is used in conjunction with a blockchain. According to~\cite{jansen2019smart}, from a theoretical point of view, a blockchain has infinite length, so there are still exist possibilities that RIDE can use Data Transactions to achieve Turing-completeness. The Turing-completeness of a blockchain system can be achieved through unwinding the recursive calls between multiple transactions and blocks instead of using a single one, and it is not necessary to have loops and recursion in the language itself~\cite{Plutus}. Generally, the idea is to consider the blockchain mechanism that generates new blocks as an infinite loop. Thus it can be used with conditions to provide the same functionality of a while-loop, allowing while-computable calculations and thus Turing-completeness.

\subsection{ Threats to validity}

\textbf{Internal Validity:} Two threats can affect the internal validity of this survey. The first is finding the related studies, which was done by searching the gray-and-white literature. Furthermore,  the search was specific to a set of sources that we decided for the white literature. Because smart contracts are an extremely diverse topic, there is a chance that we may have excluded studies or research sources that have studies related to this survey. In order to minimize this threat, we started our search by performing a general inquiry to get possibly related studies and research sources. After that, we applied our proposed search methodology with the proposed inclusion and exclusion criteria. Another threat can be the selection of the keywords in the pruning stages. To avoid having very restrictive keywords, we first tested them on previously known publications that we know from prior work. \\
\textbf{External Validity:} For the external threats to validity, we can only claim the results related to SCLs characteristics reported in our proposed classification framework but nothing outside it, although some SCLs are realized as extensions to known GPLs.

\section{Summary of Findings}
This section presents the main findings from answering our RQs and summarizes the observations of the present work. 

1.  The SCLs we reviewed are at various levels of maturity. Many are still in early development (i.e., 6 out of 36). A few (i.e., 5 out of 36) are in their alpha or beta development phase. Correspondingly, 17 languages of the total number of surveyed languages are not yet mature enough to be used in production. This leaves developers with few languages (i.e., 7 out of 36) ready to develop smart contracts.   

2.  In general, the characteristics of current SCLs help developers design and implement smart contracts. However, as a group, the languages are poor at presenting the states of smart contracts in a way that helps developers understand them prior to deploying them to the blockchain. Also, underdeveloped characteristics clarify smart contracts for regular, non-technical users by facilitating easier representations of their code. Only one of the 36 SCLs we reviewed provides users with graphical representations of contracts. 
 
3. There is limited interoperability between SCLs due to their diverse syntax. The unique syntax of each SCL continually subjects developers to new learning curve challenges. Two out of 36 SCLs (5\%) strive to facilitate the translation of contracts from one SCL to another. We see a need to develop more languages that can abstract the currently available languages and generate models or meta-models that enable transformations between these languages. This would also help in developing intermediate representations of SCLs. Developers could use these to leverage the broad spread of characteristics across many SCLs. This would be valuable considering the difficulty of having all these characteristics covered in one single language.  

4. A limited number of the surveyed SCLs (8\%) are designed with an emphasis on smart contract security. Despite the fact that smart contracts are of high economic value and inherently targets for serious attacks, SCL developers still give more attention to the usability and modularity of their language than to security. This leaves the responsibility for implementing secure contracts fully to developers, which makes the instruments more risky.

5. Whereas most of the reviewed SCLs share similar implementation goals, there is no guarantee that they share identical semantics since most of them rely on informal semantics. Given that SCLs target different platforms, this raises the possibility of semantic mismatches that affect smart contracts’ behavior during execution when more than one SCL is used to run contracts on a specific platform. Similarly, contracts written in one SCL may behave differently depending on the platform they run on (e.g., executing Solidity on different blockchains).

6. Published content that discusses SCLs is limited and does not typically provide the detailed language specifications that language engineers need. Collecting sufficient information about SCLs and their specifications forces users (and researchers) to search through research publications, code repositories, and sometimes the source files of the language itself. The field would benefit from more detailed and focused high-quality research to support language engineers and developers.
\section{Directions for Future Research}
We suggest four directions for future research, based on our review of existing Smart Contract Languages. 
\\

\textbf{First, simulating smart contract's functionality and behavior before deploying it to the blockchain.} It would be interesting and valuable to get a full analysis of an implemented smart contract using the same SCL that is used to implement it. This might help predict the contracts' non-functional properties, such as their cost, performance, and states, before the actual deployment is carried out. Also, predicting the computational costs of certain functions could help optimization engineers. Running such simulations would depend on formal definitions of the semantics and the operational semantics of the SCLs. Running such simulations would depend on formal definitions of the semantics and the operational semantics of the SCLs. Addressing this challenge can significantly enhance the feasibility and efficiency of simulations, ensuring their practical utility.

\textbf{Second, obtaining static and dynamic models for smart contracts to improve portability.} Besides static models, it would be helpful to model the operational dynamic run-time features of smart contracts. These models could capture the states of the contract, function complexity, and the smart contract account's values during run time. They could be extended to help forecast the work involved in contract upgrading, reuse, and integration. Additionally, emphasizing the potential benefits of these models in terms of enhancing portability and facilitating seamless integration with existing blockchain infrastructures would underscore their practical significance.

\textbf{Third, the use of modeling languages and meta-models to find common ground between SCLs.}
Our review of a large set of characteristics and features for 36 SCLs found many differences across languages. This remains a very heterogeneous field. We lack accepted SCL modeling concepts that would let us define commonalities across languages to allow comparison at a componential level. More mature modeling concepts would help achieve interoperability among SCLs. By establishing common modeling concepts, researchers can bridge the existing gaps and pave the way for standardized approaches to understanding and comparing SCLs. Emphasizing the collaborative and interdisciplinary nature of such standardization efforts can foster a sense of community and shared purpose among researchers.

\textbf{Fourth, embedding security methods within the current available SCLs to consolidate trust.}  
The security of smart contracts is crucial due to the amount of money that they hold. This technology's future depends on adding new security guarantees to the current SCLs or developing new methods of securing smart contracts. More efforts are needed to integrate a more advanced approach for smart contracts' verification and validation, more practices and design patterns need to be highlighted, and more security-by-design approaches need to be adopted.  Addressing these challenges requires not only highlighting the need for enhanced security but also emphasizing the broader societal impact. By ensuring the trustworthiness of smart contracts, these technologies can underpin essential services, financial transactions, and legal agreements, contributing to the overall integrity of blockchain ecosystems.

\section{Conclusion}
The rapid proliferation of smart contracts has led to the emergence of a multitude of Smart Contract Languages (SCLs) with diverse characteristics catering to developers and users. In response to this diversity, we proposed a robust classification and comparison framework. Our framework serves a dual purpose: assisting smart contract developers, researchers, and users in selecting suitable SCLs tailored to their specific needs and scrutinizing the scope, characteristics, and limitations of existing SCLs. Our comprehensive analysis identified 36 distinct SCLs, encompassing both industrial and academic domains. Notably, most of these languages are in active development, with a few achieving stability. Each of the surveyed languages benefits from an active support community, enhancing their usability and adaptability. Our examination revealed a variety of objectives, including safety, security, adaptability, and readability, each pursued to varying degrees across the languages. Syntax diversity was influenced by popular general-purpose languages, leading to a mix of text-based syntax and informal semantics. These languages span a range of abstraction levels, from high to low, with ongoing debates surrounding the need for Turing completeness. Most of the languages reviewed adopt a Domain-Specific Language (DSL) approach, with some incorporating multi-paradigm elements. Each language offers distinct programming components, enticing users and fostering ongoing language development. The intricate interplay of these characteristics underscores the dynamic landscape of SCLs, necessitating a nuanced understanding for informed decision-making and continued advancements in this field.
\subsection{Limitations}
At the time of writing (June 2020), the SCLs reviewed in this study are at various maturity levels. Even for those that are fully mature, this is still an active field. New SCLs, new updates, new features, and new software characteristics will continue to disrupt the sector. In the future, we anticipate a need to extend, update, and refine our proposed framework; to accommodate, for instance, updates of the SCLs’ features and structural programming paradigms. Our work can also be extended to encompass diverse SCL applications with more focus on user needs.


%
%


 \bibliographystyle{spmpsci}      

\bibliography{mybib.bib}

\begin{thebibliography}{10}
\providecommand{\url}[1]{{#1}}
\providecommand{\urlprefix}{URL }
\expandafter\ifx\csname urlstyle\endcsname\relax
  \providecommand{\doi}[1]{DOI~\discretionary{}{}{}#1}\else
  \providecommand{\doi}{DOI~\discretionary{}{}{}\begingroup \urlstyle{rm}\Url}\fi

\bibitem{anderson2016new}
Anderson, L., Holz, R., Ponomarev, A., Rimba, P., Weber, I.: New kids on the block: an analysis of modern blockchains.
\newblock arXiv preprint arXiv:1606.06530  (2016)

\bibitem{Simvolio}
Apla: Apla public blockchain platform for building digital ecosystems.
\newblock Available at https://apla.readthedocs.io/en/latest/topics/script.html.  (2019)

\bibitem{atzei2017survey}
Atzei, N., Bartoletti, M., Cimoli, T.: A survey of attacks on ethereum smart contracts (sok).
\newblock In: International conference on principles of security and trust, pp. 164--186. Springer (2017)

\bibitem{atzei2019developing}
Atzei, N., Bartoletti, M., Lande, S., Yoshida, N., Zunino, R.: Developing secure bitcoin contracts with bitml.
\newblock In: Proceedings of the 2019 27th ACM Joint Meeting on European Software Engineering Conference and Symposium on the Foundations of Software Engineering, pp. 1124--1128 (2019)

\bibitem{bartoletti}
Bartoletti, M., Pompianu, L.: An empirical analysis of smart contracts: platforms, applications, and design patterns.
\newblock In: International conference on financial cryptography and data security, pp. 494--509. Springer (2017)

\bibitem{biryukov2017findel}
Biryukov, A., Khovratovich, D., Tikhomirov, S.: Findel: Secure derivative contracts for ethereum.
\newblock In: International Conference on Financial Cryptography and Data Security, pp. 453--467. Springer (2017)

\bibitem{Script}
Bitcoin: Bitcoin script.
\newblock Available at https://en.bitcoin.it/wiki/Script\#Stack  (2019)

\bibitem{briscoe2011introduction}
Briscoe, C.T.: Introduction to formal semantics for natural language  (2011)

\bibitem{brooks2007fundamentals}
Brooks, D.R.: Fundamentals of the javascript language.
\newblock An Introduction to HTML and JavaScript: for Scientists and Engineers pp. 67--106 (2007)

\bibitem{Pyramid}
Burge, M.: Pyramid scheme.
\newblock Available at https://github.com/MichaelBurge/pyramid-scheme.  (2018)

\bibitem{ethereum}
Buterin, V., et~al.: Ethereum: A next-generation smart contract and decentralized application platform.
\newblock URL https://github. com/ethereum/wiki/wiki/\% 5BEnglish\% 5D-White-Paper \textbf{7} (2014)

\bibitem{h1}
Carter, G., White, D., Nalla, A., Shahriar, H., Sneha, S.: Toward application of blockchain for improved health records management and patient care.
\newblock Blockchain in Healthcare Today \textbf{3}, 10--30953 (2019)

\bibitem{coblenz2017obsidian}
Coblenz, M.: Obsidian: a safer blockchain programming language.
\newblock In: 2017 IEEE/ACM 39th International Conference on Software Engineering Companion (ICSE-C), pp. 97--99. IEEE (2017)

\bibitem{solidity}
Dannen, C.: Introducing Ethereum and solidity, vol.~1.
\newblock Springer (2017)

\bibitem{delmolino2015programmer}
Delmolino, K., Arnett, M., Kosba, A., Miller, A., Shi, E.: A programmer’s guide to ethereum and serpent.
\newblock URL: https://mc2-umd. github. io/ethereumlab/docs/serpent\_tutorial. pdf.(2015).(Accessed May 06, 2016) pp. 22--23 (2015)

\bibitem{f1}
Demirhan, H.: Effective taxation system by blockchain technology.
\newblock In: Blockchain Economics and Financial Market Innovation, pp. 347--360. Springer (2019)

\bibitem{counterparty}
Dermody, R., Krellenstein, A., Slama, O., Wagner, E.: Counterparty: protocol specification (2014) (2016)

\bibitem{LLL}
Edgington, B.: Lll compiler documentation.
\newblock Available at https://media.readthedocs.org/pdf/lll-docs/latest/lll-docs.pdf.  (2017)

\bibitem{ERights}
ERights: Erights.
\newblock Available at http:/erights.org/  (2019)

\bibitem{ewasm}
Ethereum: Ethereum webassembly (ewasm).
\newblock Available at https://ewasm.readthedocs.io/en/mkdocs/  (2019)

\bibitem{Solidity2}
Ethereum: Solidity contract language.
\newblock Available at https://solidity.readthedocs.io/  (2019)

\bibitem{Viper}
Ethereum: Viper documentation.
\newblock Available at https://vyper.readthedocs.io/en/v0.1.0-beta.13/release-notes.html  (2019)

\bibitem{Yul}
Ethereum: Yul documentation.
\newblock Available at https://solidity.readthedocs.io/en/v0.5.13/yul.html  (2019)

\bibitem{eykholt2017rchain}
Eykholt, E., Meredith, L.G., Denman, J.: Rchain architecture documentation  (2017)

\bibitem{FirmoLang}
Firmo: The future of crypto economy, firmo network - firmolang.
\newblock Available at https://steemit.com/crypto/@sjdsr/the-future-of-crypto-economy-firmo-network-firmolang  (2019)

\bibitem{Qtum}
Foundation, Q.: The qtum foundation mobile smart contracts platform.
\newblock Available at https://www.prnewswire.com/news-releases/the-qtum-foundation-releases-whitepaper-detailing-mobile-smart-contracts-platform-300419218.html.  (2018)

\bibitem{adico}
Frantz, C.K., Nowostawski, M.: From institutions to code: Towards automated generation of smart contracts.
\newblock In: 2016 IEEE 1st International Workshops on Foundations and Applications of Self* Systems (FAS* W), pp. 210--215. IEEE (2016)

\bibitem{DAO2}
Ghaleb, B., Al-Dubai, A., Ekonomou, E., Qasem, M., Romdhani, I., Mackenzie, L.: Addressing the dao insider attack in rpl’s internet of things networks.
\newblock IEEE Communications Letters \textbf{23}(1), 68--71 (2018)

\bibitem{goodman2014tezos}
Goodman, L.: Tezos: A self-amending crypto-ledger position paper.
\newblock Aug \textbf{3}, 2014 (2014)

\bibitem{goodmanimichelson}
Goodmani, L.: Michelson: the language of smart contracts in tezos.
\newblock Available at https://tezos. com/pages/tech. html  (2019)

\bibitem{grishchenko2018semantic}
Grishchenko, I., Maffei, M., Schneidewind, C.: A semantic framework for the security analysis of ethereum smart contracts.
\newblock In: International Conference on Principles of Security and Trust, pp. 243--269. Springer (2018)

\bibitem{lisk}
G{\"u}nther, S.: Lisk-the mafia blockchain.
\newblock Medium  (2018)

\bibitem{harz}
Harz, D., Knottenbelt, W.: Towards safer smart contracts: A survey of languages and verification methods.
\newblock arXiv preprint arXiv:1809.09805  (2018)

\bibitem{hirai2017defining}
Hirai, Y.: Defining the ethereum virtual machine for interactive theorem provers.
\newblock In: International Conference on Financial Cryptography and Data Security, pp. 520--535. Springer (2017)

\bibitem{Ivy-Lang}
Ivy-Lang: Ivy-lang language.
\newblock Available at https://github.com/ivy-lang/ivy-bitcoin.  (2019)

\bibitem{jansen2019smart}
Jansen, M., Hdhili, F., Gouiaa, R., Qasem, Z.: Do smart contract languages need to be turing complete?
\newblock In: International Congress on Blockchain and Applications, pp. 19--26. Springer (2019)

\bibitem{e1}
Jiang, Y., Wang, C., Wang, Y., Gao, L.: A privacy-preserving e-commerce system based on the blockchain technology.
\newblock In: 2019 IEEE International Workshop on Blockchain Oriented Software Engineering (IWBOSE), pp. 50--55. IEEE (2019)

\bibitem{jiao2018executable}
Jiao, J., Kan, S., Lin, S.W., Sanan, D., Liu, Y., Sun, J.: Executable operational semantics of solidity.
\newblock arXiv preprint arXiv:1804.01295  (2018)

\bibitem{kasampalis2019iele}
Kasampalis, T., Guth, D., Moore, B., Șerb{\u{a}}nuț{\u{a}}, T.F., Zhang, Y., Filaretti, D., Șerb{\u{a}}nuț{\u{a}}, V., Johnson, R., Ro{\c{s}}u, G.: Iele: A rigorously designed language and tool ecosystem for the blockchain.
\newblock In: International Symposium on Formal Methods, pp. 593--610. Springer (2019)

\bibitem{king2012peercoin}
King, S., Nadal, S.: Peercoin--secure \& sustainable cryptocoin.
\newblock Aug-2012 [Online]. Available: https://peercoin. net/whitepaper ()  (2012)

\bibitem{Legalese}
Legalese: L4 language.
\newblock Available at https://legalese.github.io/.  (2019)

\bibitem{namecoin}
Loibl, A., Naab, J.: Namecoin.
\newblock namecoin. info  (2014)

\bibitem{lll2}
LoomNetwork: Lll.
\newblock Available at https://github.com/loomnetwork/solidityx-jsl  (2019)

\bibitem{Solidityx}
LoomNetwork: Solidityx contract language.
\newblock Available at https://solidityx.org  (2019)

\bibitem{stellar}
Mazieres, D.: The stellar consensus protocol: A federated model for internet-level consensus.
\newblock Stellar Development Foundation \textbf{32} (2015)

\bibitem{modi2018solidity}
Modi, R.: Solidity Programming Essentials: A beginner's guide to build smart contracts for Ethereum and blockchain.
\newblock Packt Publishing Ltd (2018)

\bibitem{moonad}
Moonad: Formality, an lightweight proof-gramming language.
\newblock Available at https://github.com/moonad/Formality  (2019)

\bibitem{bitcoin}
Nakamoto, S.: Bitcoin: A peer-to-peer electronic cash system.
\newblock Tech. rep., Manubot (2019)

\bibitem{OCamlPro}
OCamlPro: The liquidity language for smart contracts.
\newblock Available at http://www.liquidity-lang.org/.  (2019)

\bibitem{o2017simplicity}
O'Connor, R.: Simplicity: A new language for blockchains.
\newblock In: Proceedings of the 2017 Workshop on Programming Languages and Analysis for Security, pp. 107--120 (2017)

\bibitem{ogawa1991towards}
Ogawa, R.T., Malen, B.: Towards rigor in reviews of multivocal literatures: Applying the exploratory case study method.
\newblock Review of educational research \textbf{61}(3), 265--286 (1991)

\bibitem{oliva2020exploratory}
Oliva, G.A., Hassan, A.E., Jiang, Z.M.J.: An exploratory study of smart contracts in the ethereum blockchain platform.
\newblock Empirical Software Engineering pp. 1--41 (2020)

\bibitem{parity1}
Palladino, S.: The parity wallet hack explained.
\newblock July-2017.[Online]. Available: https://blog. zeppelin. solutions/on-the-parity-wallet-multisig-hack-405a8c12e8f7  (2017)

\bibitem{parizit}
Parizi, R.M., Dehghantanha, A., et~al.: Smart contract programming languages on blockchains: an empirical evaluation of usability and security.
\newblock In: International Conference on Blockchain, pp. 75--91. Springer (2018)

\bibitem{pilkington2016blockchain}
Pilkington, M.: Blockchain technology: principles and applications. research handbook on digital transformations, edited by f. xavier olleros and majlinda zhegu (2016)

\bibitem{Pirapira}
Pirapira: Bamboo smart contract language.
\newblock Available at https://github.com/pirapira/bamboo  (2019)

\bibitem{daml}
Platform, T.D.A.: Daml.
\newblock Available at https://hub.digitalasset.com  (2019)

\bibitem{Plutus}
Plutus: Plutus contract language.
\newblock Available at https://github.com/input-output-hk/plutus.  (2019)

\bibitem{prieto2019blockchain}
Prieto, J., Das, A.K., Ferretti, S., Pinto, A., Corchado, J.M.: Blockchain and Applications: International Congress, vol. 1010.
\newblock Springer (2019)

\bibitem{Raineorshine}
Raineorshine: Functional solidity language.
\newblock Available at https://github.com/raineorshine/functional-solidity-language.  (2019)

\bibitem{babbage}
Reitwiessner, C.: “babbage: a mechanical smart contract language.
\newblock Online blog post  (2017)

\bibitem{Logikon}
Schaeff: Logikon, an experimental language for smart contracts.
\newblock Available at https://github.com/logikon-lang/logikon.  (2018)

\bibitem{schrans2018writing}
Schrans, F., Eisenbach, S., Drossopoulou, S.: Writing safe smart contracts in flint.
\newblock In: Conference Companion of the 2nd International Conference on Art, Science, and Engineering of Programming, pp. 218--219 (2018)

\bibitem{seijas2016scripting}
Seijas, P.L., Thompson, S.J., McAdams, D.: Scripting smart contracts for distributed ledger technology.
\newblock IACR Cryptol. ePrint Arch. \textbf{2016}, 1156 (2016)

\bibitem{sergey2018scilla}
Sergey, I., Kumar, A., Hobor, A.: Scilla: a smart contract intermediate-level language.
\newblock arXiv preprint arXiv:1801.00687  (2018)

\bibitem{szabo1996smart}
Szabo, N.: Smart contracts: building blocks for digital markets.
\newblock EXTROPY: The Journal of Transhumanist Thought,(16) \textbf{18}(2) (1996)

\bibitem{Waves}
Waves: Get smarter about what matters to you  (2019).
\newblock Available at https://blog.wavesplatform.com/waves-smart-contracts-what-to-expect-and-when-489563a95ca3

\bibitem{wohlin2014guidelines}
Wohlin, C.: Guidelines for snowballing in systematic literature studies and a replication in software engineering.
\newblock In: Proceedings of the 18th international conference on evaluation and assessment in software engineering, pp. 1--10 (2014)

\bibitem{yang2018lolisa}
Yang, Z., Lei, H.: Lolisa: Formal syntax and semantics for a subset of the solidity programming language.
\newblock arXiv preprint arXiv:1803.09885  (2018)

\end{thebibliography}
\end{document}